\documentclass[oneside,english,latin9]{aa}
\usepackage[T2A,T1]{fontenc}
\usepackage[latin9]{inputenc}
\usepackage{array}
\usepackage{multirow}
\usepackage{amsmath}
\usepackage{graphicx}
\usepackage{natbib}
\usepackage{color}

\def\pddt#1#2{\frac{\partial #1}{\partial #2}}

\makeatletter

\AtBeginDocument{\DeclareFontEncoding{T2A}{}{}}

\providecommand{\tabularnewline}{\\}

\makeatother

\usepackage{babel}
\begin{document}

\title{The effect of accretion on the pre-main-sequence evolution of low-mass stars and brown dwarfs}

\author{Eduard I. Vorobyov \inst{1,2,3}, Vardan Elbakyan \inst{2}, Takashi Hosokawa \inst{4}, 
Yuya Sakurai\inst{5}, Manuel Guedel \inst{3} and Harold Yorke \inst{6}}

\institute{Institute of Fluid Mechanics and Heat Transfer, TU Wien, Vienna, 1060, Austria;
\email{eduard.vorobiev@univie.ac.at} \and Research Institute of
Physics, Southern Federal University, Stachki Ave. 194, Rostov-on-Don,
344090, Russia;  \and University of Vienna, Department of Astrophysics, Vienna, 1180, Austria; \and Department of Physics, Kyoto University, Sakyo-ku, Kyoto,
606-8502, Japan; \and Department of Physics, Graduate School of Science, The University of Tokyo, 7-3-1 Hongo, Bunkyo-ku, Tokyo 113-0033, Japan
\and Jet Propulsion Laboratory, California Institute of Technology, Pasadena, CA 91109, USA;}

\abstract
{}
{The pre-main-sequence evolution of low-mass stars and  brown dwarfs is studied numerically
starting from the formation of a protostellar/proto-brown dwarf seed and taking 
into account the mass accretion onto the central object
during the initial several Myr of evolution.}
{The stellar evolution was computed using the STELLAR evolution code developed by 
Yorke \& Bodenheimer with recent modifications by Hosokawa et al. The mass accretion rates 
were taken from numerical hydrodynamics models of Vorobyov
\& Basu computing the circumstellar disk evolution starting from the gravitational collapse of pre-stellar cloud cores of various mass and angular momentum. The resulting stellar evolution tracks were compared with the isochrones and isomasses calculated using non-accreting models.}
{We find that mass accretion in the initial several Myr of protostellar evolution can 
have a strong effect on the subsequent evolution of young stars and brown dwarfs.  The disagreement
between accreting and non-accreting models in terms of the total stellar luminosity $L_\ast$, 
stellar radius $R_\ast$ and effective temperature $T_{\rm eff}$ depends on the 
thermal efficiency of accretion, i.e., on the
fraction of accretion energy absorbed by the central object. The largest mismatch  
is found for the cold accretion case, in which essentially all accretion energy is radiated away. 
The relative deviations in $L_\ast$ and $R_\ast$ in this case can reach 50\% for 1.0-Myr-old
objects and remain notable even for 10-Myr-old objects.
In the hot and hybrid accretion cases, in which a constant fraction of accretion energy is absorbed, the disagreement between accreting and non-accreting models becomes less pronounced, but still remains
notable for 1.0-Myr-old objects. These disagreements may lead to the wrong age estimate for objects
of (sub-)solar mass when using the isochrones based on non-accreting models, as was also previously
noted by Baraffe et al. and Hosokawa et al. We find that objects with
strong luminosity bursts exhibit notable excursions in the $L_\ast$--$T_{\rm eff}$ diagram, but the
character of these excursions is distinct for hybrid/hot and cold accretion scenarios. In particular,
the cold accretion scenario predicts peak luminosities that are greater than those of known FU-Orionis-type
outbursts, which implies that cold accretion is physically less realistic.}
{Mass accretion during the early stages of star and brown dwarf evolution is an important factor, but
its effect depends on the details of how accretion energy is distributed within the star.}


\authorrunning{E. Vorobyov et al.}
\titlerunning{The effect of accretion on stars and brown dwarfs}

\maketitle

\section{Introduction}

The evolution of the pre-main-sequence (PMS) stars has been studied for decades by various authors 
\citep[e.g.,][]{Baraffe1998,Palla2000}. Regardless of some model uncertainties, 
stellar evolution calculations show the standard evolution path that goes through Hayashi
and Henyey tracks on the Herzsprung-Russell (HR) diagram in consensus \citep{Henyey1955,Hayashi1961}.
Effects of the mass accretion have been also incorporated in stellar evolution calculations 
\citep[e.g.,][]{Stahler1980,Palla1991,Hosokawa2009}. 
\citet{Palla1990} propose the concept of the protostellar birth line, where the PMS 
stars first becomes optically visible as the surrounding envelope disperses and the mass 
accretion ceases.

Recent studies further add updates considering the mass accretion in more realistic situations.
First of all, while most of the previous studies assume simple constant accretion histories for 
simplicity, numerical simulations are revealing a more sporadic nature of the mass accretion. 
Since parental molecular cores have finite amounts of angular momentum, a protostar accretes 
the gas through a circumstellar disk which forms thanks to the near-conservation of the angular momentum
in the collapsing core. In the early embedded phase of disk evolution where most of 
the final stellar mass is accumulated, the angular momentum and mass transport is mostly driven by the gravitational torque \citep{VB2009b}. In this phase, the disks are often prone to gravitational instability,
which results in very time-dependent mass accretion histories, e.g., the episodic 
accretion histories repeating short accretion burst events and relatively longer quiescent phases
\citep{VB2010,VB2015,Machida2011,Tomida2017}.
Other mechanisms, such as the magneto-rotational and thermal instabilities, planet-disk interactions,
and close stellar encounters can also produce episodic accretion bursts 
both in the embedded and T Tauri phases of disk evolution 
\citep[see][for a review]{Audard2014}. There are also various 
observational signatures of such episodic accretion reported for low-mass protostars
\citep[e.g.,][]{Dunham2010,Liu2016}.

Effects of such variable mass accretion are recently included in stellar evolution calculations 
focusing on the early  \citep{Vorobyov2016,Hosokawa2016} and late evolutionary phases
\citep[e.g.,][]{Baraffe2009,Baraffe2012,Baraffe2016,Hosokawa2011}.
These studies present that PMS tracks can largely differ from the non-accreting tracks that 
have been used to estimate the stellar age spreads in young clusters \citep[e.g.,][]{Soderblom2014}.
In some cases, a PMS star first appears fairly near the main-sequence on the HR diagram 
immediately after the mass accretion ceases, while the non-accreting isochrones assume 
that the PMS star starts the evolution in an upper part on the HR diagram. 
As a result, the age of the star as inferred from the non-accreting tracks can be overestimated
by about a factor of 2.

Although the above mismatch was found in the studies with the variable 
mass accretion, there is another factor incorporated in the model normally with a free-parameter: 
the thermal efficiency of the accretion, i.e., the specific entropy of the accreted gas.
It is actually suggested that such ``warmth'' of the accretion controls the PMS evolution 
rather than the time-variability of the accretion rates  
\citep[e.g.,][]{Hartmann1997,Hartmann2016,Hosokawa2011,Kunitomo2017}.
However, these effects are often mixed because the accretion
thermal efficiency can be modeled as a function of the accretion rates \citep{Baraffe2012}. 
It is still quantitatively 
uncertain to what extent the non-accreting isochrones are reliable after all, and 
whether the accretion variability or thermal efficiency of the accretion plays the major role 
in the PMS evolution.

In this work, we revisit the PMS evolution of (sub-)solar mass stars and brown dwarfs 
taking variable mass accretion into consideration and making different 
assumptions regarding the thermal efficiency parameter $\alpha$, the fraction of the accretion 
energy taken into
the stellar interior. We use variable mass accretion histories on a protostar ($\dot{M}$) obtained 
in numerical simulations following the formation and evolution of protostars and protostellar
disks described in detail in \citet{VB2010}. Then we compute the evolution of the 
protostars using a modified version of the stellar evolution code STELLAR 
\citep{Yorke2008,Hosokawa2013,Sakurai2015}.  This approach is similar to that employed 
in \citet{Baraffe2012}, but using the STELLAR code rather than the Lyon code \citep{Baraffe2009}.
We compare the PMS evolutionary tracks with the above variable
accretion to the non-accreting isochrones calculated with the same code STELLAR, and monitor how 
large offsets the accreting tracks show with the isochrones and isomasses as functions of the time.
Moreover, we also quantitatively analyze how such evolution of the offsets varies with different 
modeling of the thermal efficiency $\alpha$. Our results show that, with $\alpha > 10^{-3}$
(we call hybrid and hot accretion cases below), offsets with the isochrones and isomasses only appears for 
early ages of $<$ a few Myr, where even the non-accreting isochrones suffers from relatively 
large modeling uncertainties. At later ages, the notable offsets remain only if 
$\alpha$ is fixed at $10^{-3}$ (we call cold accretion below), 
clearly showing that how small $\alpha$ is allowed is a key to alter the PMS tracks.

The rest of the paper is organized as follows. We first explain our adopted methods on 
the numerical simulations of the disk accretion and on the stellar evolution calculations
in Sections 2 and 3 respectively. The obtained results are described in Sections 4-6, where 
the cases are divided in terms of the different thermal efficiencies, i.e., the hybrid, hot, 
and cold accretion. In Section 7, we propose possible updates of the isochrones
considering the effects of the mass accretion. Finally Sections 8 and 9 are devoted 
to the discussions and conclusions.

\section{Model description}

We compute the evolution of low-mass stars and upper-mass brown dwarfs%
\footnote{ Hereafter, we refer to both as stars and differentiate between stars
and brown dwarfs only when it is explicitly needed.%
} using the stellar evolution code STELLAR originally developed by
\citet{Yorke2008} with recent modifications described in Section~\ref{stellar}.
We start our computations from a protostellar seed, continue through
the main accretion phase where the growing star accumulates most of
its final mass, and end our computations when the star approaches
the main sequence. For the protostellar accretion rates, we use those
obtained from the numerical hydrodynamics simulations of prestellar
core collapse described in Section \ref{sub:Numerical-hydrodynamics-code}.
We note that this approach is not fully self-consistent. The stellar
evolution calculations are not fully coupled with the disk evolution,
but employ the precomputed mass accretion rates. The full coupling
of disk evolution modeling with stellar evolution calculations  was
recently done in \citet{Baraffe2016} using the Lyon code. 
Such real-time coupling has also been performed with the STELLAR code,
mostly for studying the high-mass star formation
\citep[e.g.,][]{KY13,Hosokawa2016}.
We also plan to apply STELLAR code for future studies on the
low-mass star formation.

\subsection{Numerical hydrodynamics code\label{sub:Numerical-hydrodynamics-code}}

The evolution of protostellar disks is computed using the numerical
hydrodynamics code described in detail in \citet{VB2010}. Here, we
briefly review the main concepts and equations. We start our numerical
simulations from the gravitational collapse of a gravitationally contracting
cloud core, continue into the embedded phase of star formation, during
which a star, disk, and envelope are formed, and terminate our simulations
after 1.0--2.0 Myr of evolution, depending on the model and available computational 
resources\footnote{Models with higher stellar masses are usually more resource 
demanding due to shorter time steps.}. 
The protostellar disk, when
formed, occupies the inner part of the numerical grid, while the collapsing
envelope occupies the rest of the grid. As a result, the disk is not
isolated but is exposed to intense mass loading from the envelope in the embedded phase
of star formation.
To avoid too small time steps, we introduce a ``sink cell\textquotedblright{}
near the coordinate origin with a radius of $r_{\mathrm{sc}}$ = 5
AU and impose a free outflow boundary condition so that the matter
is allowed to flow out of the computational domain, but is prevented
from flowing in. The mass accretion rate is calculated as the mass
passing through the sink cell per time step of numerical integration.
We assume that 90\% of the gas that crosses the sink cell lands onto
the protostar. A small fraction of this mass (a few per cent) remains
in the sink cell to guarantee a smooth transition of the gas surface
density across the inner boundary. The other 10\% of the accreted
gas is assumed to be carried away with protostellar jets.

The main physical processes taken into account when computing the
evolution of the disk and envelope include viscous and shock heating,
irradiation by the forming star, background irradiation, radiative
cooling from the disk surface and self-gravity. The code is written
in the thin-disk limit, complemented by a calculation of the disk
vertical scale height using an assumption of local hydrostatic equilibrium.
The resulting model has a flared structure, which guaranties that
both the disk and envelope receive a fraction of the irradiation energy
from the central protostar. The corresponding equations of mass, momentum,
and energy transport are

\begin{equation}
\frac{\partial\Sigma}{\partial t}=-\nabla_{\mathrm{p}}\cdot\left(\Sigma v_{\mathrm{p}}\right),\label{eq:mass}
\end{equation}

\begin{equation}
\frac{\partial}{\partial t}\left(\Sigma v_{\mathrm{p}}\right)+\left[\nabla\cdot\left(\Sigma v_{\mathrm{p}}\otimes v_{\mathrm{p}}\right)\right]_{\mathrm{p}}=-\nabla_{\mathrm{p}}P+\Sigma g_{\mathrm{p}}+\left(\nabla\cdot\Pi\right)_{\mathrm{p}},\label{eq:momentum}
\end{equation}

\begin{equation}
\frac{\partial e}{\partial t}+\nabla_{\mathrm{p}}\cdot\left(ev_{\mathrm{p}}\right)=-P\left(\nabla_{\mathrm{p}}\cdot v_{\mathrm{p}}\right)-\varLambda+\varGamma+\left(\nabla\cdot v\right)_{\mathrm{pp'}}:\varPi_{\mathrm{pp'}},\label{eq:energy}
\end{equation}
where subscripts $p$ and $p'$ refers to the planar components $(r,\phi)$
in polar coordinates, $\Sigma$ is the mass surface density, $e$
is the internal energy per surface area, $P$ is the vertically integrated
gas pressure calculated via the ideal equation of state as $P=(\gamma-1)e$
with $\gamma=7/5$, $v_{p}=v_{\mathrm{r}}\hat{r}+v_{\mathrm{\phi}}\hat{\phi}$
is the velocity in the disk plane, and $\nabla_{\mathrm{p}}=\hat{r}\partial/\partial r+\hat{\phi}r^{-1}\partial/\partial\phi$
is the gradient along the planar coordinates of the disk. The gravitational
acceleration in the disk plane, $g_{\mathrm{p}}=g_{\mathrm{r}}\hat{r}+g_{\mathrm{\phi}}\hat{\phi}$,
takes into account self-gravity of the disk, found by solving for
the Poisson integral (see details in \citet{VB2010}), and the gravity
of the central protostar when formed. Turbulent viscosity due to sources
other than gravity is taken into account via the viscous stress tensor
$\varPi$, the expression for which is provided in \citet{VB2010}.
We parameterize the magnitude of kinematic viscosity $\nu$ using
the alpha prescription with a spatially and temporally uniform $\alpha_{\mathrm{visk}}=5\times10^{-3}$.

The radiative cooling $\varLambda$ in equation (\ref{eq:energy})
is determined using the diffusion approximation of the vertical radiation
transport in a one-zone model of the vertical disk structure

\begin{equation}
\varLambda=F_{\mathrm{c}}\sigma T_{\mathrm{mp}}^{4}\frac{\tau}{1+\tau^{2}},
\end{equation}
where $\sigma$ is the Stefan-Boltzmann constant, $T_{\mathrm{mp}}=P\mu/R\Sigma$
is the midplane temperature of gas, $\mu$ = 2.33 is the mean molecular
weight (90\% of atomic H, 10\% of atomic He), \textit{R} is the universal
gas constant, and $F_{\mathrm{c}}=2+20\mathrm{tan^{-1}}(\tau)/(3\pi)$
is a function that secures a correct transition between the optically
thick and optically thin regimes. We use frequency-integrated opacities
of Bell \& Lin (1994). The heating function is expressed as

\begin{equation}
\Gamma=F_{\mathrm{c}}\sigma T_{\mathrm{irr}}^{4}\frac{\tau}{1+\tau^{2}},
\end{equation}
where $T_{\mathrm{irr}}$ is the irradiation temperature at the disk
surface determined by the stellar and background black-body irradiation
as

\begin{equation}
T_{\mathrm{irr}}^{4}=T_{\mathrm{bg}}^{4}+\frac{F_{\mathrm{irr}}(r)}{\sigma},
\end{equation}
where $T_{\mathrm{bg}}$ is the uniform background temperature (in
our model set to the initial temperature of the natal cloud core)
and $F_{\mathrm{irr}}(r)$ is the radiation flux (energy per unit
time per unit surface area) absorbed by the disk surface at radial
distance \textit{r} from the central star. The latter quantity is
calculated as

\begin{equation}
F_{\mathrm{irr}}(r)=\frac{L_{\ast}}{4\pi r^{2}}cos\gamma_{\mathrm{irr}}
\end{equation}
where $L_{\ast}$ is the total (accretion plus photospheric) luminosity
of the central object (star or brown dwarf), $\gamma_{\mathrm{irr}}$ is the incidence angle of radiation
arriving at the disk surface at a radial distance \textit{r}. The
incidence angle is calculated using the disk surface curvature inferred
from the radial profile of the disk vertical scale height (see \citet{VB2010},
for more details). The photospheric luminosity $L_{\rm ph}$ is calculated from the D'Antona
\& Mazzitelly stellar evolution tracks \citep{DAM}, while the accretion luminosity 
is calculated as follows:
\begin{equation}
L_{\rm accr} = \epsilon {G M_\ast \dot{M} \over R_\ast},
\end{equation}
where  
$M_\ast$ the mass of the central object, and $R_\ast$ the radius of the central object (also provided
by the D'Antona \& Mazzitelly tracks).  We note that D'Antona \& Mazitelli's tracks do not 
cover the very early phases of stellar evolution. Therefore, we have used a
power-law expression to extrapolate to times earlier than those included in the pre-main-sequence 
tracks, $L_{\rm ph} = L_{\rm ph,0} (t/t0)^4$, where $t_0$ is the earliest time in the tracks 
and $L_{\rm ph,0}$ is the pre-main-sequence luminosity at this time. 
We assume $\epsilon=1/2$, which is characteristic of accretion from a thin disk. Because we use the non-accretion isochrones when calculating the mass accretion
histories, the 
thermal efficiency of accretion $\alpha$ cannot be taken into account. However,
in the stellar evolution calculations using the pre-computed accretion histories, we do 
take the thermal efficiency of accretion  into account (see Section~\ref{stellar}.
Equations (\ref{eq:mass})-(\ref{eq:energy}) are solved on the polar grid $(r,\phi)$ with a 
numerical resolution of $512\times512$ grid
points. The numerical procedure is described in detail in \citet{VB2010}.

\subsection{Initial conditions for the numerical hydrodynamics code}

The initial radial profiles of the gas surface density $\Sigma$ and
angular velocity $\Omega$ of collapsing pre-stellar cores have the
following form: 
\begin{equation}
\Sigma=\frac{r_{0}\Sigma_{0}}{\sqrt{r^{2}+r_{0}^{2}}}\label{eq:sigma}
\end{equation}
\begin{equation}
\Omega=2\Omega_{0}\left(\frac{r_{0}}{r}\right)^{2}\left[\sqrt{1+\left(\frac{r}{r_{0}}\right)^{2}}-1\right]\label{eq:omega}
\end{equation}
Here, $\Sigma_{0}$ and $\Omega_{0}$ are the gas surface density and angular velocity 
at the center of the core and $r_{0}=\sqrt{A}c_{\mathrm{s}}^{2}/\pi G\Sigma_{0}$
is the radius of the central plateau, where $c_{\mathrm{s}}$ is the
initial sound speed in the core. These radial profiles are typical
of pre-stellar cores formed as a result of the slow expulsion of magnetic
field due to ambipolar diffusion, with the angular momentum remaining
constant during axially-symmetric core compression (Basu 1997). The
value of the positive density perturbation \textit{A} is set to 1.2
and the initial gas temperature in collapsing cores is set to 10 K.

Each model is characterized by a distinct ratio $r_{\mathrm{out}}/r_{0}=6$
in order to generate gravitationally unstable truncated cores of similar
form, where $r_{\mathrm{out}}$ is the cloud core's outer radius.
The actual procedure for generating the parameters of a specific core
consists in two steps. First, we choose the outer cloud core radius
$r_{\mathrm{out}}$ and find $r_{0}$ using the adopted ratio between
these two quantities. Then, we find the central surface density $\Sigma_{0}$
from the relation $r_{0}=\sqrt{A}c_{\mathrm{s}}^{2}/\pi G\Sigma_{0}$
and determine the resulting cloud core mass $M_{\mathrm{cl}}$ from
Equation (\ref{eq:sigma}). Once the gas surface density profile and
the initial mass are fixed, we set the angular velocity profile by
choosing the value of $\Omega_{0}$ so that the ratio of rotational
to gravitational energy $\beta$ falls within the limit inferred by
Caselli et al. (2002) for dense molecular cloud cores. The parameters
of our models are presented in Table~\ref{tab:1}. 
The second column is the initial core mass, the third column the ratio
of rotational to gravitational energy, the fourth column the radius
of the central near-constant-density plateau, and the fifth column
is the final stellar mass. The models are
ordered in the sequence of increasing final stellar masses $M_{*,\,\mathrm{fin}}$.

\subsection{Accretion rate histories}

\label{accretion}

The available computational resources allow us to
calculate the disk evolution and, hence, the protostellar accretion
history only for about 1.0--2.0 Myr. We assume that in the subsequent
evolution the mass accretion rate declines linearly to zero during
another 1.0 Myr. Effectively, this implies a disk dispersal time of
1 Myr and the total disk age of about 2.0--3.0 Myr in our models. These values are
in general agreement with the disk ages inferred from observations
(Williams \& Cieza 2011), though some objects demonstrate longer disk
lifetimes.

\begin{table}
\begin{centering}
\protect\protect\caption{\label{tab:1}Initial parameters for all 35 models used
in the current work.}
\begin{tabular}{ccccc}
\hline 
Model  & $M_{\mathrm{core}}$  & $\beta$  & $r_{0}$  & $M_{*,\,\mathrm{fin}}$\tabularnewline
 & {[}$M_{\odot}${]}  & {[}\%{]}  & {[}AU{]}  & {[}$M_{\odot}${]}\tabularnewline
\hline 
\hline 
1  & 0.061  & 11.85  & 137  & 0.042\tabularnewline
2  & 0.077  & 8.72  & 171  & 0.052\tabularnewline
3  & 0.085  & 2.23  & 189  & 0.078\tabularnewline
4  & 0.099  & 2.24  & 223  & 0.089\tabularnewline
5  & 0.092  & 0.57  & 206  & 0.090\tabularnewline
6  & 0.108  & 1.26  & 240  & 0.103\tabularnewline
7  & 0.123  & 2.24  & 274  & 0.107\tabularnewline
8  & 0.122  & 0.57  & 274  & 0.120\tabularnewline
9  & 0.154  & 2.24  & 343  & 0.128\tabularnewline
10  & 0.154  & 1.26  & 343  & 0.141\tabularnewline
11  & 0.200  & 0.56  & 446  & 0.194\tabularnewline
12  & 0.230  & 1.26  & 514  & 0.201\tabularnewline
13  & 0.307  & 0.56  & 686  & 0.273\tabularnewline
14  & 0.384  & 1.26  & 857  & 0.307\tabularnewline
15  & 0.461  & 2.25  & 1029  & 0.322\tabularnewline
16  & 0.538  & 1.26  & 1200  & 0.363\tabularnewline
17  & 0.461  & 0.56  & 1029  & 0.392\tabularnewline
18  & 0.430  & 0.28  & 960  & 0.409\tabularnewline
19  & 0.615  & 0.56  & 1372  & 0.501\tabularnewline
20  & 0.692  & 1.27  & 1543  & 0.504\tabularnewline
21  & 0.584  & 0.28  & 1303  & 0.530\tabularnewline
22  & 0.845  & 2.25  & 1886  & 0.559\tabularnewline
23  & 0.922  & 1.27  & 2057  & 0.579\tabularnewline
24  & 0.738  & 0.28  & 1646  & 0.643\tabularnewline
25  & 0.845  & 0.56  & 1886  & 0.653\tabularnewline
26  & 1.245  & 1.27  & 2777  & 0.753\tabularnewline
27  & 1.076  & 0.56  & 2400  & 0.801\tabularnewline
28  & 1.767  & 2.25  & 3943  & 0.807\tabularnewline
29  & 0.999  & 0.28  & 2229  & 0.818\tabularnewline
30  & 1.537  & 1.27  & 3429  & 0.887\tabularnewline
31  & 1.306  & 0.28  & 2915  & 1.031\tabularnewline
32  & 1.383  & 0.56  & 3086  & 1.070\tabularnewline
33  & 1.844  & 1.27  & 4115  & 1.100\tabularnewline
34  & 1.767  & 0.28  & 3943  & 1.281\tabularnewline
35  & 1.690  & 0.56  & 3772  & 1.322\tabularnewline
\hline 
\end{tabular}
\par\end{centering}
\end{table}

\begin{figure*}
\begin{centering}
\includegraphics[scale=0.9]{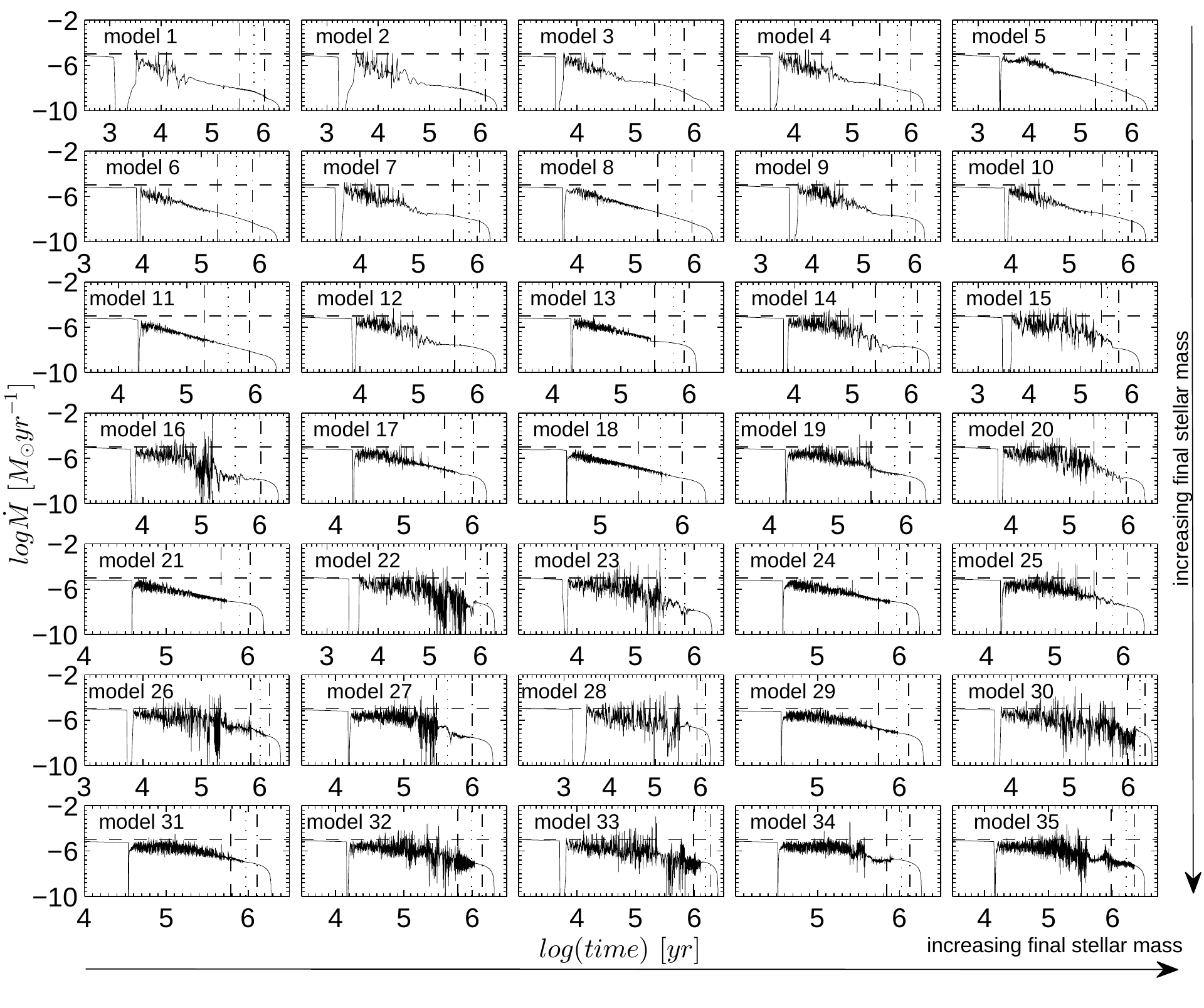} 
\par\end{centering}
\protect\protect\caption{\label{fig:1}Mass accretion rates vs. time for 35 models. The horizontal
dashed lines mark the critical value for the transition from cold
to hot accretion scenario.  The models are arranged so that the final stellar mass increases
from left to right and from top to bottom. The vertical dashed, dotted, and dash-dotted lines 
mark time instances when  90\%, 95\%, and 98\% of the total stellar mass has been accumulated.}
\end{figure*}

We calculate protostellar accretion histories for 35 pre-stellar cores
listed in Table \ref{tab:1}. Figure \ref{fig:1} presents $\dot{M}$
vs. time elapsed since the beginning of numerical simulations. The horizontal dashed lines mark a
critical value of the mass accretion rate $\dot{M}_{{\rm cr}}=10^{-5}~M_\odot$~yr$^{-1}$ above
which accretion changes from cold to hot in the hybrid accretion scenario (see Section~\ref{Ascenarios}).
The protostellar accretion rate is calculated as the mass passing
through the sink cell per unit time, $\dot{M}=-2\pi r_{{\rm sc}}\Sigma v_{{\rm r}}$,
where $v_{r}$ is the radial component of velocity at the inner computational
boundary.

Evidently, the accretion rates in the low-$M_{\mathrm{core}}$ and
low-$\beta$ models (e.g., models 8, 11, 13, 18, 21) are characterized
by an order-of-magnitude flickering, which gradually diminishes with
time. On the other hand, models with high $M_{\mathrm{core}}$ and
$\beta$ (e.g., models 14, 16, 22, 23) demonstrate large-amplitude
variations in $\dot{M}$ and strong accretion bursts exceeding in
magnitude $10^{-5}~M_{\odot}$~yr$^{-1}$. This difference in the
time behaviour of $\dot{M}$ stems from the different properties of
protostellar disks formed from the gravitational collapse of prestellar
cores \citep{Vorobyov2010}. The low-$M_{\mathrm{core}}$ and low-$\beta$
models produce disks of low mass and size, which are weakly gravitationally
unstable and show no sign of fragmentation, while high $M_{\mathrm{core}}$
and $\beta$ models form disks that are sufficiently massive and extended
to develop strong gravitational instability and fragmentation. The
forming fragments often migrate onto the star owing to the loss of
angular momentum via gravitational interaction with spiral arms or
other fragments in the disk, producing strong accretion bursts similar
in magnitude to FU-Orionis-type eruptions \citep{VB2006,VB2010,VB2015}.

\begin{figure*}
\begin{centering}
\includegraphics[scale=1.1]{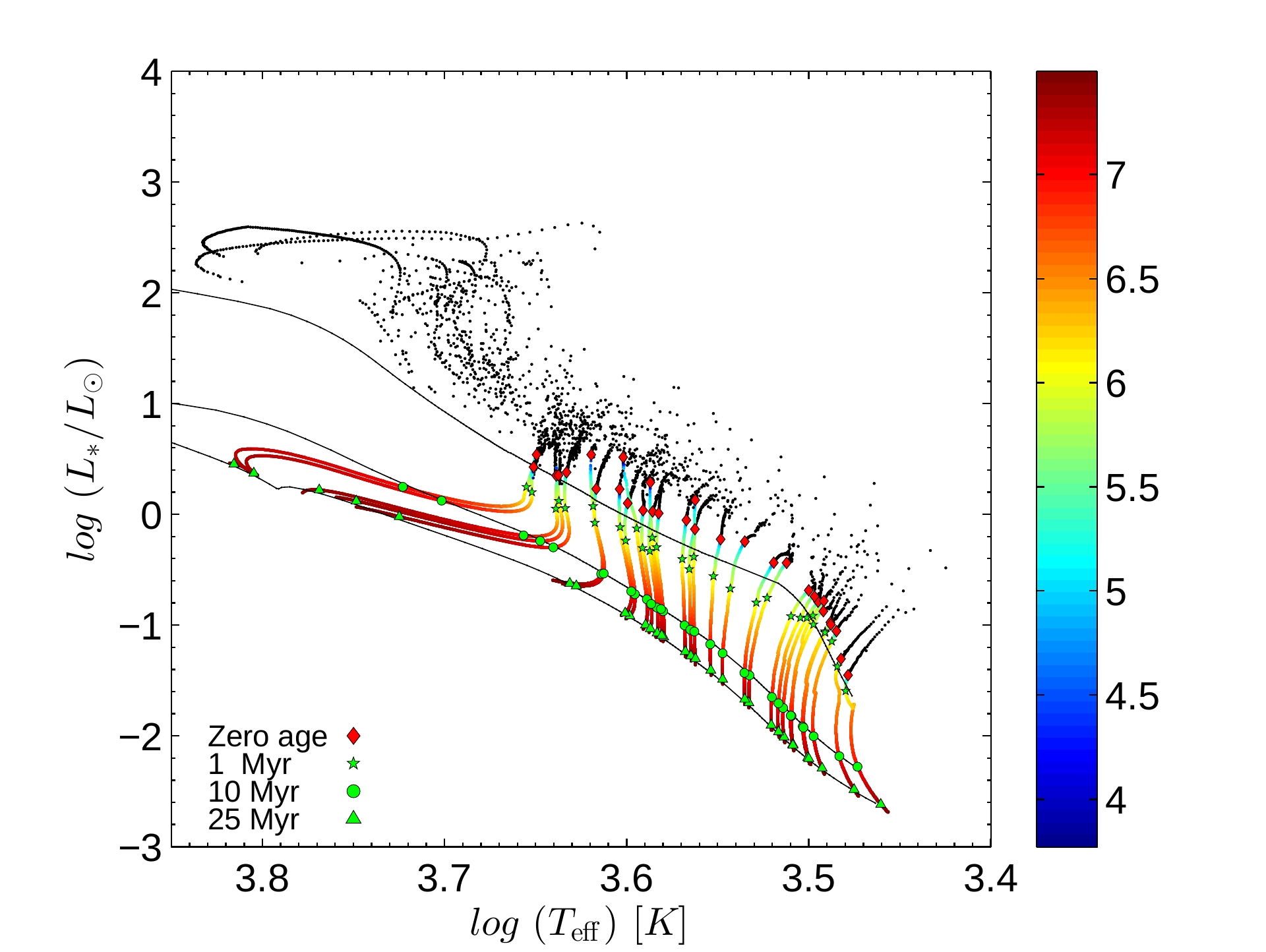} 
\par\end{centering}
\protect\protect\caption{\label{fig:2} Stellar evolution sequences on 
the total luminosity $L_\ast$ -- effective temperature $T_{\mathrm{eff}}$
diagram for the hybrid accretion models. The dots present the model
tracks and the color of the dots is varying according to the stellar
age shown in the vertical bar (in log yr). The zero point age for each model is
marked by the red diamonds. The green symbols mark the reference
ages as indicated in the bottom-left corner, elapsed since the zero point age of each
object. The black solid
lines provide the isochrones for stellar ages of 1 Myr, 10 Myr and
25 Myr (from top to bottom) derived from non-accreting stellar evolution
models of \citet{Yorke2008}.}
\end{figure*}

\section{Stellar evolution code}

\label{stellar} We use the stellar evolution code STELLAR originally
developed by Yorke \& Bodenheimer (2008). The detailed description
of the code can be found in Sakurai et al. (2015). We here briefly
review the main features of the code. 

The basic equations to be solved are as follows:
\begin{align}
\pddt{r}{m}=&\frac{1}{4\pi r^2\rho}, \\[2 pt]
\pddt{P}{m}=&-\frac{Gm}{4\pi r^4}, \\[2 pt]
\pddt{l}{m}=&E_{\rm nuc}-c_{P}\pddt{T}{t}+\frac{\delta}{\rho}\pddt{P}{t}, \\[2 pt]
\pddt{T}{m}=&-\frac{GmT}{4\pi r^4P}\nabla,
\end{align}
where $m$ is the mass contained within a spherical layer with radius $r$, 
$P$ the total (gas plus radiation) pressure, 
$l$ the local luminosity, $E_{\rm nut}$ the specific energy production rate by nuclear reactions, 
$c_P$ the isobaric specific heat, $T$ the temperature,
$\delta\equiv -(\partial\ln\rho/\partial\ln T)_P$, and $\nabla\equiv\partial\ln T/\partial\ln P$
the temperature gradient calculated using the mixing-length theory
for convective layers.
Nuclear reactions are computed up to the helium burning ($3\alpha$ and \{CNO\}$~+$He).

The model stars are divided into two parts: the atmosphere, which provides surface boundary conditions, and the interior, which contains most of the stellar material.  We consider the atmosphere to be  spherically symmetric and gray, applying the equations of hydrostatic equilibrium and radiative/convective energy transport, which are integrated inward by a Runge-Kutta method.  We integrate the atmospheric structure down to a fitting point, where the boundary conditions for the interior are provided. We use the standard Henyey method to get the interior structure converged to satisfy the boundary conditions. As the accreting gas settles onto the stellar surface, the added material gradually sinks inward through the atmosphere and is eventually incorporated into the interior by a rezoning procedure.

\subsection{The initial conditions for the stellar evolution code}

The stellar evolution calculations start from a fully convective,
polytropic stellar seed of 5 Jupiter masses and 3 Jovian radii with
a polytropic index $n=1.5$. Before commencing actual calculations
taking into account the mass growth via accretion, the polytropic
seed is allowed to relax to a fully converged stellar model. 

\subsection{The thermal efficiency of accretion}
\label{Ascenarios}
Once the properties of the initial protostellar seed are set, we commence
the stellar evolution calculations using the accretion rate histories
calculated in Section~\ref{accretion}. During these calculations,
we assume that a fraction $\alpha$ of the accretion energy $\epsilon GM_{\ast}\dot{M}/R_{\ast}$
is absorbed by the protostar, while a fraction $1-\alpha$ is radiated
away and contributes to the accretion luminosity of the star. Here,
$M_{\ast}$ and $R_{\ast}$ are the mass and radius of the central
star.  In this paper, we consider three scenarios for the thermal efficiency of accretion: \textcolor{blue}{(i)}
cold accretion with a constant $\alpha=10^{-3}$, meaning that practically
all accretion energy is radiated away and little is absorbed by the
star,  \textcolor{blue}{(ii)} hot accretion with a constant $\alpha=0.1$, and 
\textcolor{blue}{(iii)} a hybrid scheme defined as follows:
\begin{equation}
\alpha  = \left\{ \begin{array}{ll} 
   10^{-3},  &\,\,\,  \mbox{if $\dot{M}<10^{-7}~M_\odot$~yr$^{-1}$ }, \\ 
   \dot{M}\times10^4 \left[{\mathrm{yr} \over M_\odot} \right],  & \,\,\, 
   \mbox{if $10^{-7}~M_\odot~\mathrm{yr}^{-1} \le  \dot{M}  \le 10^{-5}~M_\odot$~yr$^{-1}$} ,  \\
   0.1,  & \,\,\, \mbox{if $\dot{M}>10^{-5}~M_\odot$~yr$^{-1}$}.
   \end{array} 
   \right. 
   \label{function} 
\end{equation}
This functional form of $\alpha$ has a property that accretion remains cold
at small $\dot{M}$ and gradually changes to hot accretion above 
a certain critical value of $\dot{M}_{\mathrm{cr}}$, for which we chose 
$10^{-5}M_{\odot}$~yr$^{-1}$ based on modeling of FU-Orionis-type eruptions 
of \citet{Kley1996} and 
\citet{Hartmann2011} and analytical calculations of \citet{Baraffe2012}
showing that such a transition reflects a change in the accretion
geometry with increasing accretion rate and/or in the magnetospheric
interaction between the star and the disk. We note that the actual
value of $\alpha$ in the hybrid accretion scheme is changing smoothly
from $10^{-3}$ to 0.1 over a time period of $\sim1.0$~Myr to preserve
numerical stability of the stellar evolution code.

\section{Hybrid accretion}
\label{hybridaccr}
In this section, we present our model results for the hybrid accretion
scenario, in which the value of $\alpha$ depends on the mass accretion
rate. The computed stellar evolution sequences of the total (accretion plus photospheric) luminosity
$L_\ast$ vs. effective temperature $T_{\mathrm{eff}}$ for 31
models\footnote{We have excluded models 15, 26, 28, 30 because of the numerical problems with the stellar
evolution code.} 
are shown with the coloured dots in Figure \ref{fig:2}. The color of the dots varies according
to the stellar age indicated on the vertical bar (in log yr). 
The zero point of the stellar age is defined as the time instance when the growing star 
accumulates 95\% of its final mass. 
At this time instance, the envelope is essentially dissipated and the disk-to-star mass ratio
is 5\%. This value is in agreement with the inferred upper limit on the disk-to-star mass ratios 
of young T~Tauri stars in nearby star-forming regions 
\citep[e.g.][]{Pascucci2016}, especially considering 
that the disk masses may in fact be somewhat underestimated when using dust continuum measurements 
\citep{Dunham2014,Tsukamoto2016}. 
In other words, the adopted 95\% criterion for the zero age corresponds
to young T Tauri stars that have just left the protostellar phase of stellar evolution. We want also
to emphasize the qualitative change in character of mass accretion at this stage (see Figure~\ref{fig:1})
-- $\dot{M}$ is smooth and declining after the stars accumulate 95\% of  
the final mass,
whereas in the earlier evolution period $\dot{M}$ can be quite variable 
featuring strong episodic bursts.  
The effect of variations in this quantity is discussed later in the text. 
This time instance is marked for every model in Figure~\ref{fig:2}
by the red diamonds. The black dots correspond to the evolutionary phase 
preceding the zero point age for each object\footnote{These data points are shown 
only starting from the moment of disk formation, since the evolutionary tracks before disk formation
are almost identical because of the identical initial temperature of the pre-stellar cores.}. 
We will refer to this phase as the protostellar phase of evolution. 
The green symbols mark the reference ages of 1~Myr,
10~Myr, and 25~Myr for each model. The black solid lines present the
isochrones for the same ages, but derived from the non-accreting stellar
evolution models of \citet{Yorke2008} (hereafter,
the non-accreting isochrones).



\begin{figure*}
\begin{centering}
\includegraphics[scale=0.6]{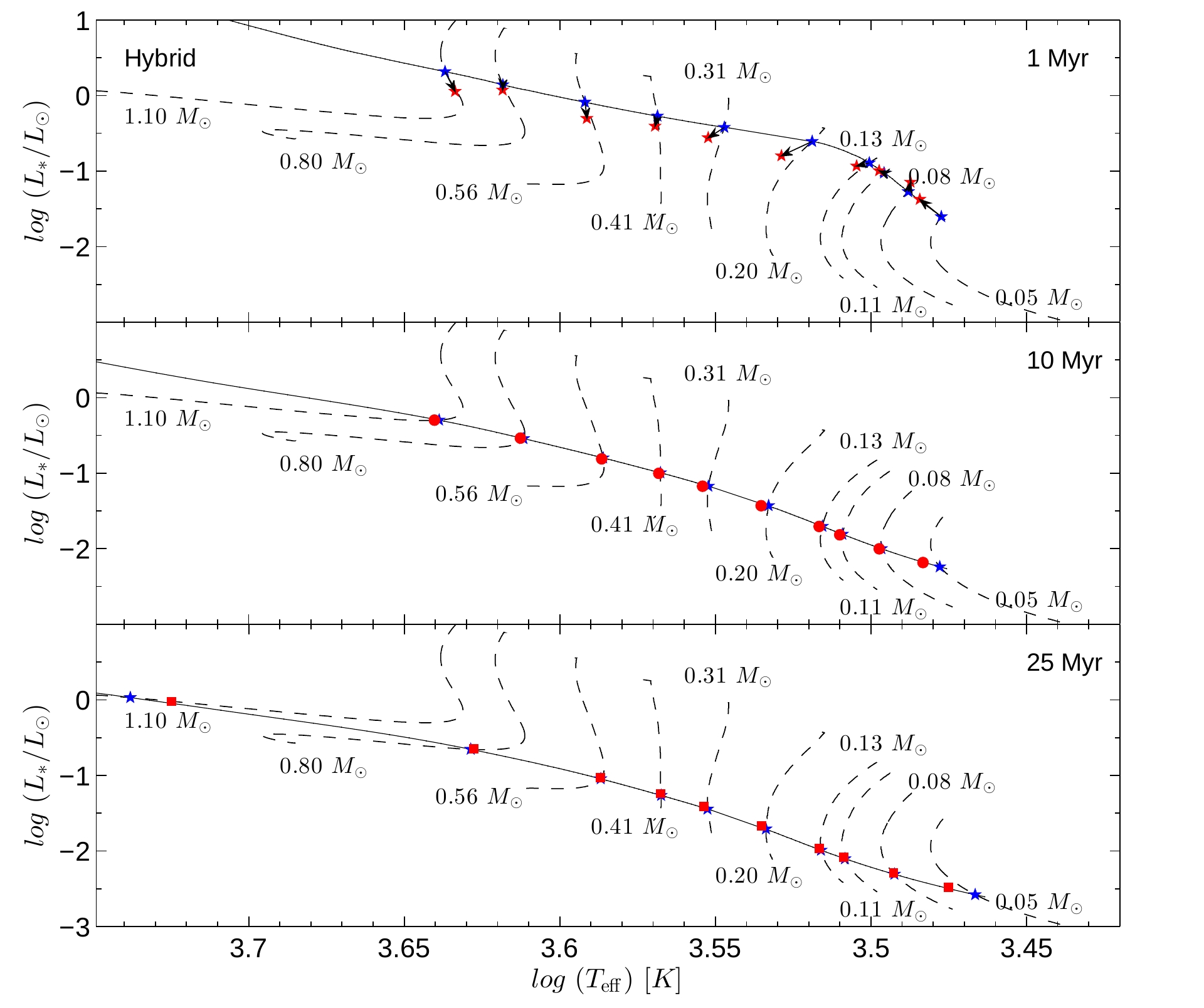} 
\par\end{centering}

\protect\protect\caption{\label{fig:3}HR diagrams for models with the age for 1 Myr (top subplot),
10 Myr (center) and 25 Myr (bottom). The black solid lines indicate
non-accreting isochrones for 1 Myr (top subplot), 10 Myr (center)
and 25 Myr (bottom). The black dashed lines show the non-accreting
isomasses. Red and blue markers indicate accreting and non-accreting
models, respectively, at the age of 1 Myr (stars), 10 Myr (circles),
25 Myr (squares) and with a mass (in solar masses) shown next to each
isomass they fit. Black arrows are showing the difference between
accreting and non-accreting models with the same mass and age.}
\end{figure*}

Evidently, young objects occupy the upper region of the $L_{\ast}$--$T_{\mathrm{eff}}$
diagram, shifting toward the bottom-left part as they age. Notable
excursions of young protostellar objects (black dots) to the upper-left corner of the diagram
are caused by strong mass accretion bursts discussed in Section~\ref{accretion}.
During these bursts both the stellar luminosity and effective temperature increase.
The contraction of the star after the accretion burst occurs on the thermal relaxation timescale
$t_{\rm therm}$, which can be defined  in our case as 
\begin{equation}
t_{\rm therm}=\frac{1}{\langle{L_{\rm ph}} \rangle} \sum_{i}\alpha_{i}\frac{G M_{\ast,i}\dot{M_{i}}}{R_{\ast,i}}
\Delta t_{i},
\label{relax}
\end{equation}
where the summation is performed over the burst duration defined as a time interval 
when $\alpha_i\ne0$ (or $\dot{M}>10^{-5}~M_\odot$~yr$^{-1}$). The expression under
the sum gives the total  accretion energy absorbed by the star during the burst and this
energy is divided by the mean stellar photospheric luminosity averaged over the duration of the relaxation
period. We found that the stars recover approximately the previous equilibrium on the thermal relaxation
timescales.
Our analysis of $t_{\rm therm}$ shows that the star can spend from hundreds to
up to ten thousand of years (depending on the burst strength)
in the peculiar excursion tracks, where it can potentially be confused with more massive stars
in quiescence. A more detailed study of this phenomenon is deferred for a follow-up paper.

A visual inspection of Figure~\ref{fig:2} indicates that young objects
show a notable deviation from the non-accreting isochrones. 
This is especially evident for the models with effective temperatures 
$\log T_{\rm eff} \ge 3.5$, which lie notably lower than the 1.0-Myr-old isochrone,
meaning that these objects appear older on the $L_\ast$--$T_{\rm eff}$ diagram 
than they truly are\footnote{The effective temperature of $10^{3.5}\approx3160$~K on the 
1.0-Myr-old non-accreting isochrone corresponds to a star with mass $M_{\ast}=0.13~M_{\odot}$.}. 
Our results agree with findings of \citet{Hosokawa2011},
which state that the non-accreting isochrones can sometimes overestimate
stellar ages for stars with effective temperatures above $\ge3500$~K
($\log T_{{\rm eff}}\approx3.54$). Our results agree in general with numerical
models of \citet{Baraffe2016}, but their low-mass objects usually show 
stronger deviations from the non-accreting isochrones. 

The non-accreting isochrones are
often used to derive the stellar ages using the measured bolometric
luminosities and effective temperatures. Our modeling demonstrates
that this practice needs to be taken with care, especially for young stars with
effective temperatures $\ge3200$~K, which may look older than they truly
are when using the non-accreting isochrones. 
For older stars, however, we obtained a much better fit with 
the 10-Myr-old and 25-Myr-old non-accreting isochrones.

\begin{table*}
\begin{centering}
\protect\protect\caption{\label{tab:2}Differences in luminosity ($\Delta L_\ast$), effective temperature
($\Delta T_{\mathrm{eff}}$) and stellar radius ($\Delta R_\ast$) between accreting
and non-accreting models  in the hybrid accretion scenario.}
\begin{tabular}{c||ccc|}
\multirow{1}{*}{} &  & 1 Myr & \tabularnewline
\hline 
\hline 
$M_\ast$ & $\Delta L_\ast$  & $\Delta T_{\mathrm{eff}}$  & $\Delta R_\ast$\tabularnewline
{[}$M_{\odot}${]}  & {[}\%{]}  & {[}\%{]}  & {[}\%{]}\tabularnewline
\hline 
\hline 
0.05  & 69.17 & 1.60 & 23.98\tabularnewline
0.08  & 34.11 & -0.17 & 15.35\tabularnewline
0.11  & 6.91 & 0.36 & 2.67\tabularnewline
0.13  & -9.31 & 0.94 & -6.54\tabularnewline
0.20  & -35.62 & 2.31 & -25.45\tabularnewline
0.31  & -27.09 & 1.20 & -20.31\tabularnewline
0.41  & -26.42 & 0.17 & -18.04\tabularnewline
0.56  & -39.30 & -0.14 & -21.86\tabularnewline
0.80  & -14.47 & 0.02 & -11.78\tabularnewline
1.10  & -45.25 & -0.74 & -24.90\tabularnewline
\hline 
\end{tabular}%
\begin{tabular}{ccc|}
\multicolumn{3}{c}{1 Myr (98\%)}\tabularnewline
\hline 
\hline 
$\Delta L_\ast$  & $\Delta T_{\mathrm{eff}}$  & $\Delta R_\ast$\tabularnewline
\textbf{{[}\%{]} } & \textbf{{[}\%{]} } & \textbf{{[}\%{]}}\tabularnewline
\hline 
\hline 
45.99 & 1.77 & 16.71\tabularnewline
27.15 &0.03 & 12.23\tabularnewline
-18.12 & 1.19 & -11.62\tabularnewline
-29.69 & 1.70 & -18.95\tabularnewline
-50.50 & 2.67 & -33.05\tabularnewline
-46.01 & 1.32 & -28.42\tabularnewline
-43.23 & 0.20 & -24.91\tabularnewline
-46.07 & -0.30 & -26.12\tabularnewline
-36.91 & -0.22 & -20.80\tabularnewline
-50.42 & -0.88 & -28.31\tabularnewline
\hline 
\end{tabular}%
\begin{tabular}{ccc|}
\multicolumn{3}{c|}{10 Myr}\tabularnewline
\hline 
\hline 
$\Delta L_\ast$  & $\Delta T_{\mathrm{eff}}$  & $\Delta R_\ast$\tabularnewline
{[}\%{]}  & {[}\%{]}  & {[}\%{]}\tabularnewline
\hline 
\hline 
14.03 & 1.26 & 4.22\tabularnewline
-0.27 & 0.16 & -0.46\tabularnewline
-1.69 & 0.19 & -1.22\tabularnewline
-0.60 & 0.24 & -0.77\tabularnewline
-0.09 & 0.56 & -0.55\tabularnewline
0.25  & 0.43 & -0.76\tabularnewline
-1.53 & 0.13 & -1.00\tabularnewline
-2.65 & 0.12 & -1.52\tabularnewline
-3.05 & 0.13 & -1.80\tabularnewline
4.00  & 1.41 & -0.81\tabularnewline
\hline 
\end{tabular}%
\begin{tabular}{ccc|}
\multicolumn{3}{c}{25 Myr}\tabularnewline
\hline 
\hline 
$\Delta L_\ast$  & $\Delta T_{\mathrm{eff}}$  & $\Delta R_\ast$\tabularnewline
{[}\%{]}  & {[}\%{]}  & {[}\%{]}\tabularnewline
\hline 
\hline 
24.34 & 1.99 & 7.20\tabularnewline
4.07  & 0.07 & 1.88\tabularnewline
4.67  & 0.10 & 2.10\tabularnewline
6.50  & 0.15 & 2.86\tabularnewline
9.31  & 0.33 & 4.44\tabularnewline
9.45  & 0.28 & 3.99\tabularnewline
6.25  & 0.08 & 2.92\tabularnewline
3.66  & 0.04 & 1.74\tabularnewline
0.11  & -0.34 & 0.77\tabularnewline
1.43  & -1.27 & 3.39\tabularnewline
\hline 
\end{tabular}
\par\end{centering}

\end{table*}

It is important to compare our models with the non-accreting isochrones not only for  
objects of similar age, but also for objects of similar mass.
We illustrate this  in Figure~\ref{fig:3}, where we plot
both the isochrones and {\it isomasses} derived from the non-accreting
stellar evolution models of \citet{Yorke2008}. In
particular, the solid lines in the top, middle, and bottom panels
show the non-accreting isochrones for the 1.0~Myr, 10~Myr, and 25~Myr
old stars, respectively. The dashed lines in all panels present the isomasses for
several chosen values of the stellar mass. The intersection between a specific
isochrone and a specific isomass represents a stellar or sub-stellar object 
with the corresponding age and mass. These intersections are marked with the blue symbols.
The red symbols denote objects with the same mass and age, but derived
from our numerical models. The black arrows indicate the difference
between our accreting models and non-accreting models of Yorke \& Bodenheimer for similar mass and age.
The resulting relative deviations (in per cent) in total luminosities, effective temperatures,
and stellar radii of accreting models from the non-accreting ones
are summarized in Table~\ref{tab:2} for 10 objects shown in Figure~\ref{fig:3}. 
The deviations are calculated using the following formula:
\begin{equation}
\label{difference}
\Delta X=100\frac{X_{\mathrm{accr}}-X_{\mathrm{non-accr}}}{X_{\mathrm{non-accr}}},
\end{equation}
where $X_{\rm accr}$ and $X_{\rm non-accr}$ stand for $L_\ast$, $T_{\rm eff}$, or
$R_\ast$ in accreting and non-accreting models, respectively. 
For the 1.0-Myr-old objects, we have also calculated the deviations assuming the stellar zero age 
at the time instance when protostars accumulate 98\% of their final mass (in contrast
to the originally adopted value of 95\%). This new value implies the disk-to-star mass ratio of
2\%, which is more appropriate for somewhat more evolved T~Tauri stars.

Evidently, there exist notable differences (on the order of tens of per cent)  
between the accreting and non-accreting 1.0-Myr-old models in terms of the total luminosity
for models of {\it all} masses, including the very-low-mass stars and brown dwarfs.
This cannot be easily seen from Figure~\ref{fig:2},
which in fact suggests a relatively good fit between the accreting and non-accreting models 
for objects with $\log T_{\rm eff}<3.5$. This is, however, a mere coincidence and the accreting 
and non-accreting models of the same mass are shifted along 
(but not away from) the non-accreting isochrone. 
This example illustrates the importance of using both the stellar ages and masses when
making a comparison of accreting and non-accreting models, because the accreting
objects may fall on the non-accreting isochrone, but be displaced along the isochrone 
when compared with non-accreting objects of the same mass.
All accreting models except for the least massive ones have smaller
luminosities than their non-accreting counterparts, as can be seen from the sign of the 
calculated differences in Table~\ref{tab:2}. 

A disagreement of similar magnitude
between accreting and non-accreting models is also found for the stellar radius,
while the corresponding effective temperatures are characterized by a much smaller 
deviation. On the other hand, the 10-Myr-old and 25-Myr-old accreting 
and non-accreting models show a much better agreement with each other. 
We note that the accreting models with the 98\% zero age definition 
show on average  higher deviations from the non-accreting models of the same mass and age.

Table~\ref{tab:2} provide the differences between accreting 
and non-accreting models only for 10 selected models.
In order to check if the found tendency takes place for all models in our sample, 
we present in Table~\ref{tab:2-2} the values of the differences averaged (by absolute value) 
over models with $M_{\rm\ast,fin}>0.2~M_\odot$ and models with 
$M_{\rm\ast,fin}<0.2~M_\odot$. In agreement with our previous conclusions, 
the largest deviations are found for the total luminosities and stellar radii and 
the smallest deviations are found for the effective temperatures. In general, the
deviations diminish as the stars age and the averaged differences 
for the 10-Myr-old and 25-Myr-old objects are almost
an order of magnitude lower than those for the 1.0-Myr-old ones.

\begin{table*}
\begin{centering}
\protect\protect\caption{\label{tab:2-2}Differences in luminosity ($\langle \Delta L_\ast \rangle$),
effective temperature ($\langle \Delta T_{\mathrm{eff}} \rangle$) and stellar radius 
($\langle \Delta R_\ast \rangle$) between accreting
and non-accreting models  in the hybrid accretion scenario averaged (by absolute value) 
separately over all 
models with $M_{\rm\ast,fin}>0.2~M_\odot$ and $M_{\rm\ast,fin}<0.2~M_\odot$.  }
\textbf{{}}%
\begin{tabular}{c||ccc|}
\multirow{1}{*}{} &  & {{1 Myr}} & \tabularnewline
\hline 
\hline 
{{$M_{*,\,\mathrm{fin}}$}} & {{$\langle \Delta L_\ast \rangle$ }} & {{$\langle 
\Delta T_{\mathrm{eff}}\rangle$ }} & $\langle \Delta R_\ast \rangle$ \tabularnewline
{[}$M_{\odot}${]}  & {[}\%{]}  & {[}\%{]}  & {[}\%{]}\tabularnewline
\hline 
\hline 
< 0.2 & 50.80 & 1.25 & 17.74\tabularnewline
> 0.2 & 17.40 & 0.28 & 11.00\tabularnewline
\hline 
\end{tabular}\textbf{{}}%
\begin{tabular}{ccc|}
\multicolumn{3}{c|}{10 Myr}\tabularnewline
\hline 
\hline 
$\langle \Delta L_\ast\rangle$ & $\langle\Delta T_{\mathrm{eff}}\rangle$ & $\langle\Delta R_\ast\rangle$\tabularnewline
{[}\%{]} & {[}\%{]}  & {[}\%{]}\tabularnewline
\hline 
\hline 
19.33 & 1.15 & 5.94\tabularnewline
1.51 & 0.26 & 0.52\tabularnewline
\hline 
\end{tabular}\textbf{{}}%
\begin{tabular}{ccc|}
\multicolumn{3}{c}{25 Myr}\tabularnewline
\hline 
\hline 
$\langle\Delta L_\ast\rangle$ & $\langle\Delta T_{\mathrm{eff}}\rangle$ & $\langle\Delta R_\ast\rangle$\tabularnewline
\textbf{{{[}\%{]} }} & \textbf{{{[}\%{]} }} & \textbf{{{[}\%{]}}}\tabularnewline
\hline 
\hline 
28.79  & 1.38 & 8.43\tabularnewline
2.81 & 0.29 & 1.29\tabularnewline
\hline 
\end{tabular}
\par\end{centering}

\end{table*}

\section{Hot accretion}

\begin{figure*}
\begin{centering}
\includegraphics[scale=1.1]{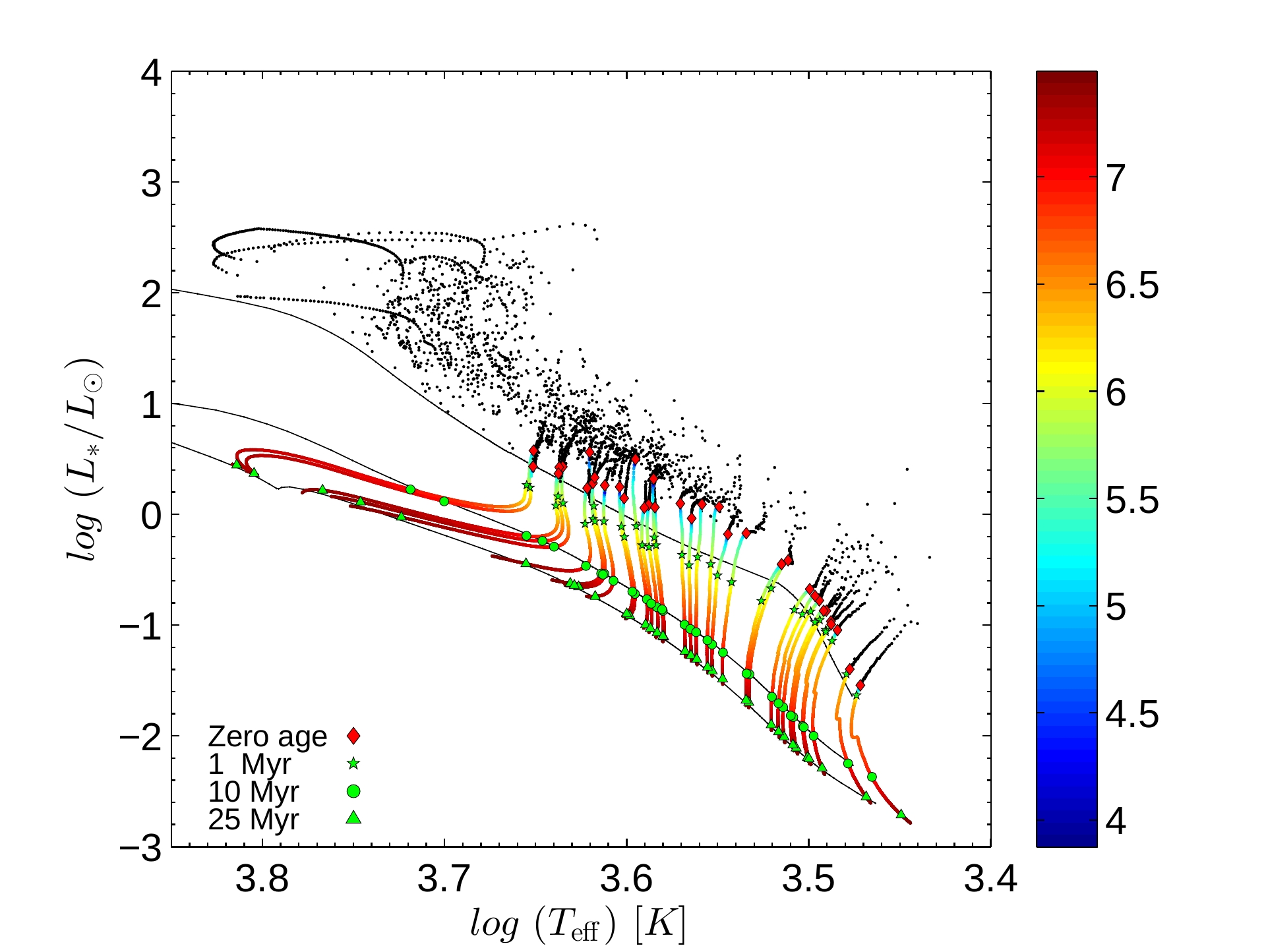} 
\par\end{centering}

\protect\protect\caption{\label{fig:4}Similar to Figure \ref{fig:2} but for the hot accretion
scenario}
\end{figure*}

In this section, we present the results for the hot accretion scenario,
in which $\alpha$ is set to 0.1 during the entire evolution period.
Figure \ref{fig:4} shows the stellar evolutionary sequences for all 35
models. The notations are the same as in Figure~\ref{fig:2}.
A visual comparison of Figures~\ref{fig:2}
and \ref{fig:4} indicates that the hot accretion models behave similarly
to the hybrid accretion ones. The 1.0-Myr-old objects
show a moderate mismatch with the corresponding 1.0-Myr-old non-accreting isochrone.
On the other hand, the 10-Myr-old and 25-Myr-old objects 
show a good agreement with the corresponding non-accreting isochrones.

As with the hybrid accretion scenario, we quantify the disagreement 
between the accreting and non-accreting models  showing in Table~\ref{tab:3}
the relative differences in the total luminosity ($\Delta L_\ast$), 
effective temperature ($\Delta T_{\rm eff}$), and stellar radius ($\Delta R_\ast$) 
for objects with the same mass and age. For the sake of comparison, 
we have chosen the same 10 objects  as in Table~\ref{tab:2}. 
There again exists a moderate deviation between the accreting and non-accreting 1.0-Myr-old
models in terms of the total luminosity and stellar radius, but this disagreement diminishes for
10-Myr-old and 25-Myr-old models. At the same time, the deviation in the effective temperature 
is rather small for all models irrespective of their mass and ages. 
All in all, the behaviour of the hybrid and hot accretion models is similar, in agreement 
with the previous work of \citet{Baraffe2012}. 
However, one difference between our work and that of Baraffe et al. is worth emphasizing:
for lowest stellar masses our accreting models at 1~Myr predict stellar luminosities and radii that
are higher than those of non-accreting isochrones (see Tables~\ref{tab:2} and \ref{tab:3}), while in Baraffe et al. the accreting models 
yield lower $L_\ast$ and $R_\ast$. We note that for stellar masses $\ge 0.15-0.2~M_\odot$, 
both works agree
and predict lower $L_\ast$ and $R_\ast$ for accreting models as compared to their non-accreting counterparts. We note that the STELLAR non-accreting isochrone shows a sudden turn towards lower luminosities at around
$T_{\rm eff}=3500$~K (corresponding to $\approx 0.15-0.2~M_\odot$) and this feature is not seen in the
Lyon non-accreting isochrones, which can partly explain the found mismatch. However, the exact reasons
requires further  investigation and comparison between the two stellar evolution codes (STELLAR and Lyon), which is beyond the scope of this work.

Finally, in Table~\ref{tab:2-2-1} we show the averaged
differences in $L_\ast$, $T_{\rm eff}$ and $R_\ast$ calculated separately for all models with 
$M_{\rm\ast,fin}>0.2~M_\odot$ and all models with 
$M_{\rm\ast,fin}<0.2~M_\odot$. Here, we also find no systematic and significant differences 
between the hybrid accretion and hot accretion models.

\begin{table*}
\begin{centering}
\protect\protect\caption{\label{tab:3}Differences in luminosity ($\Delta L_\ast$), effective temperature
($\Delta T_{\mathrm{eff}}$) and stellar radius ($\Delta R_\ast$) between accreting
and non-accreting models  in 
the hot accretion scenario.}
\begin{tabular}{c||ccc|}
\multirow{1}{*}{} &  & 1 Myr & \tabularnewline
\hline 
\hline 
$M_\ast$ & $\Delta L_\ast$ & $\Delta T_{\mathrm{eff}}$ & $\Delta R_\ast$\tabularnewline
{[}$M_{\odot}${]} & {[}\%{]} & {[}\%{]} & {[}\%{]}\tabularnewline
\hline 
\hline 
-0.05 & 45.22 & 0.43 & 17.58\tabularnewline
-0.08 & 35.58 & -0.20 &  16.12\tabularnewline
-0.11 & 12.25 & 0.17 & 5.64\tabularnewline
-0.13 & -2.23 & 0.69 & -2.51\tabularnewline
-0.20 & -33.22 & 1.67 & -22.81\tabularnewline
-0.31 & -26.14 & 0.69 & -18.79\tabularnewline
-0.41 & -19.52 & 0.23 & -13.87\tabularnewline
-0.56 & -35.74 & -0.12 & -19.63\tabularnewline
-0.80 & -13.94  & -0.08 & -11.25\tabularnewline
-1.10  & -47.61 & -0.70 & -27.71\tabularnewline
\hline 
\end{tabular}%
\begin{tabular}{ccc|}
 & 10 Myr & \tabularnewline
\hline 
\hline 
$\Delta L_\ast$ & $\Delta T_{\mathrm{eff}}$ & $\Delta R_\ast$\tabularnewline
{[}\%{]} & {[}\%{]} & {[}\%{]}\tabularnewline
\hline 
\hline 
-1.42 & 0.13  & -0.89\tabularnewline
0.07 & 0.16 & -0.28\tabularnewline
-1.17 & 0.19 & -0.95\tabularnewline
-0.20  & 0.24 & -0.54\tabularnewline
-1.30 & 0.29 & -0.64\tabularnewline
-0.89 & 0.21 & -0.83\tabularnewline
-0.18 & 0.13 & -0.36\tabularnewline
-1.78 & 0.12 & -1.10\tabularnewline
0.71 & 0.11 & 0.14\tabularnewline
3.37 & 1.05  & -0.35\tabularnewline
\hline 
\end{tabular}%
\begin{tabular}{ccc|}
 & 25 Myr & \multicolumn{1}{c}{}\tabularnewline
\hline 
\hline 
$\Delta L_\ast$ & $\Delta T_{\mathrm{eff}}$ & $\Delta R_\ast$\tabularnewline
{[}\%{]} & {[}\%{]} & {[}\%{]}\tabularnewline
\hline 
\hline 
6.60 & 0.46  & 2.39 \tabularnewline
4.01 & 0.07 & 1.88\tabularnewline
4.77 & 0.10 & 2.17\tabularnewline
6.31  & 0.15 & 2.80\tabularnewline
7.27 & 0.16 & 3.84\tabularnewline
7.94 & 0.12 & 3.65\tabularnewline
6.79 & 0.10 & 3.17\tabularnewline
3.90 & 0.01 & 1.86\tabularnewline
0.17 & -0.53 & 1.17\tabularnewline
0.82 & -1.69 & 3.97\tabularnewline
\hline 
\end{tabular}
\par\end{centering}

\end{table*}

\begin{table*}

\protect\protect\caption{{\label{tab:2-2-1}} Similar to Table~\ref{tab:2-2} but for the hot
 accretion scenario}

\begin{centering}
\textcolor{black}{}%
\begin{tabular}{c||ccc|}
\multirow{1}{*}{} &  & \textcolor{black}{1 Myr} & \tabularnewline
\hline 
\hline 
\textcolor{black}{$M_{*,\,\mathrm{fin}}$} & \textcolor{black}{$\langle\Delta L_\ast\rangle$ } & 
\textcolor{black}{$\langle \Delta T_{\rm eff}\rangle$ } & \textcolor{black}{$\langle \Delta R_\ast\rangle$}\tabularnewline
\textcolor{black}{{[}$M_{\odot}${]} } & \textcolor{black}{{[}\%{]} } & \textcolor{black}{{[}\%{]} } & \textcolor{black}{{[}\%{]}}\tabularnewline
\hline 
\hline 
\textcolor{black}{< 0.2} & \textcolor{black}{21.31 } & \textcolor{black}{0.38} & \textcolor{black}{10.15}\tabularnewline
\textcolor{black}{> 0.2} & \textcolor{black}{20.78 } & \textcolor{black}{0.27} & \textcolor{black}{12.90}\tabularnewline
\hline 
\end{tabular}\textcolor{black}{}%
\begin{tabular}{ccc|}
\multicolumn{3}{c|}{\textcolor{black}{10 Myr}}\tabularnewline
\hline 
\hline 
\textcolor{black}{$\langle\Delta L_\ast\rangle$ } & \textcolor{black}{$\langle\Delta T_{\rm eff}\rangle$ } & \textcolor{black}{$\langle\Delta R_\ast\rangle$}\tabularnewline
\textcolor{black}{{[}\%{]} } & \textcolor{black}{{[}\%{]} } & \textcolor{black}{{[}\%{]}}\tabularnewline
\hline 
\hline 
\textcolor{black}{0.55} & \textcolor{black}{0.18} & \textcolor{black}{0.49}\tabularnewline
\textcolor{black}{1.51} & \textcolor{black}{0.23} & \textcolor{black}{0.53}\tabularnewline
\hline 
\end{tabular}\textcolor{black}{}%
\begin{tabular}{ccc|}
\multicolumn{3}{c}{\textcolor{black}{25 Myr}}\tabularnewline
\hline 
\hline 
\textcolor{black}{$\langle\Delta L_\ast\rangle$ } & \textcolor{black}{$\langle\Delta T_{\rm eff}\rangle$ } & \textcolor{black}{$\langle\Delta R_\ast\rangle$}\tabularnewline
\textcolor{black}{{[}\%{]} } & \textcolor{black}{{[}\%{]} } & \textcolor{black}{{[}\%{]}}\tabularnewline
\hline 
\hline 
\textcolor{black}{5.41} & \textcolor{black}{0.13} & \textcolor{black}{2.42}\tabularnewline
\textcolor{black}{3.25} & \textcolor{black}{0.40} & \textcolor{black}{1.51}\tabularnewline
\hline 
\end{tabular}
\par\end{centering}

\end{table*}

\section{Cold accretion}

\begin{figure*}
\begin{centering}
\includegraphics[scale=1.1]{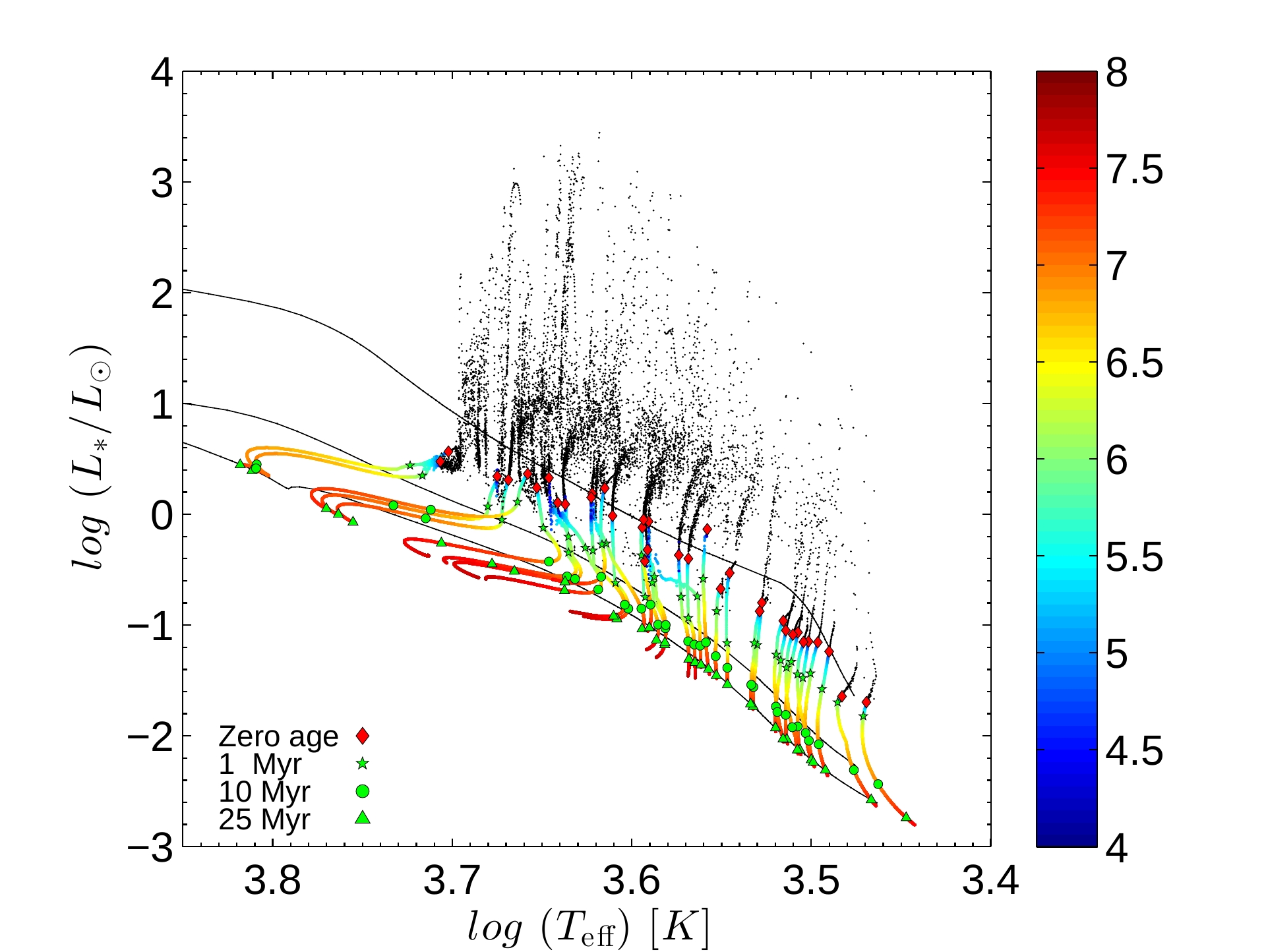} 
\par\end{centering}
\protect\protect\caption{\label{fig:6}Similar to Figure \ref{fig:2} but for the cold accretion
scenario.}
\end{figure*}

In this section, we present our results for the cold accretion scenario,
in which $\alpha$ is always set to a small value of $10^{-3}$ independent 
of the actual value of the mass accretion rate, meaning that almost all accretion energy is radiated
away and only a tiny fraction is absorbed by the protostar. 
Figure \ref{fig:6} shows the stellar evolutionary sequences for all 35 models of Table~\ref{tab:1}. The meaning of the coloured symbols and lines is the same as in Figure~\ref{fig:2}.
The disagreement between our accreting models and the non-accreting
isochrones is evident not only for the 1.0-Myr-old objects, but also for the 10-Myr-old
ones. In particular, several 1.0-Myr-old accreting objects lie near the 10-Myr-old non-accreting 
isochrone. Notable differences with the non-accreting isochrones are even present for 
the 25-Myr-old objects with $\log T_{\rm eff} > 3.55$. 

The cold accretion objects experiencing accretion bursts  
show strong surges in the luminosity, moving to the upper part of the HR diagram. 
This is due to the fact that almost all accretion energy is radiated away and the
star does not expand dramatically during the burst, having similar effective temperatures and 
radii during and between the bursts (see Figure~\ref{fig:9}). We note that the hybrid/hot accretion
objects experience an increase in both luminosity and effective temperature, moving to the upper-left
part of the HR diagram. The increase in luminosity
during the bursts is however notably stronger, sometimes exceeding 1000~$L_\odot$, than
with hybrid/hot accretion, in which case the stellar luminosity does not exceed 400~$L_\odot$.
This difference can again be explained by the compact size of the cold accretion objects 
during the bursts, which results in higher accretion luminosities as compared to
those of more bloated hybrid/hot accretion objects. According to the recent review by \citet{Audard2014},
the strongest luminosity outbursts are observed in FU~Ori (340-500~$L_\odot$), Z~Cma (400-600~$L_\odot$),
and V1057 (250-800~$L_\odot$). None of these object's luminosities exceed $1000~L_\odot$. 
We note that our burst models show a good agreement with observations in terms of the model accretion
rates \citep{VB2015}.  
Therefore, the lack of very strong luminosity outbursts in FU-Orionis-type 
objects may be an indirect evidence in favour of the hybrid/hot scenario for the thermal efficiency
of accretion.

\begin{figure*}
\begin{centering}
\includegraphics[scale=0.7]{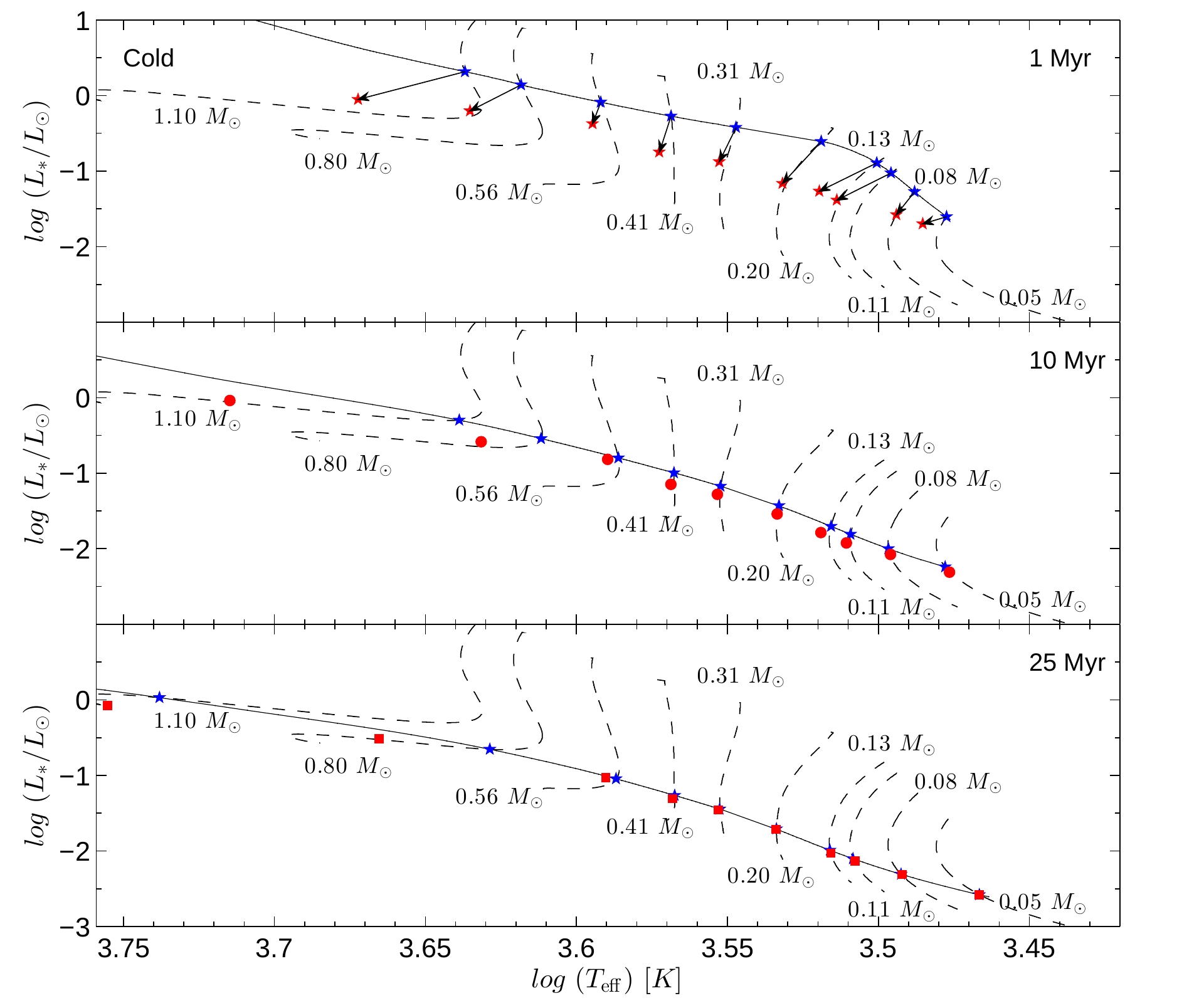} 
\par\end{centering}
\protect\protect\caption{\label{fig:7}Similar to Figure \ref{fig:3} but for the cold accretion
scenario.}
\end{figure*}

In Figure~\ref{fig:7} we compare the total luminosities and effective temperatures 
of 10 selected accreting models with the non-accreting models of Yorke \& Bodenheimer for the same mass and age. The meaning of the lines and symbols is the same as in Figure~\ref{fig:3}.
A comparison of Figures~\ref{fig:3} and \ref{fig:7} shows that the 1.0-Myr-old cold accretion
models are  characterized  by much stronger deviations from the corresponding non-accreting 
models than was found for the hybrid accretion case. 
More importantly, the cold accretion models show notable deviations at 10~Myr and, for 
upper-mass objects, even at 25 Myr.
To quantify the disagreement between the
accreting and non-accreting models, the relative deviations
in the total luminosity, effective temperature, and stellar radius for
objects with the same mass and age are calculated using Equation~(\ref{difference}) 
and are shown in Table~\ref{tab:6}. In addition, Table~\ref{tab:7} summarizes the 
averaged differences in $L_\ast$, $T_{\rm eff}$ and $R_\ast$ calculated separately 
for all models with $M_{\rm\ast,fin}>0.2~M_\odot$ and all models with 
$M_{\rm\ast,fin}<0.2~M_\odot$.

On average, the disagreement between the cold accretion models and the corresponding 
non-accreting models of Yorke \& Bodenheimer has increased
by about a factor of several as compared to the case of hybrid accretion.
More specifically, the deviations in $L_\ast$ for the 1.0-Myr-old models 
have grown by a factor of 2, while the deviations in $T_{\rm eff}$ and $R_\ast$
have grown by factors of 3--4. The disagreement for older models is even more notable: the deviation in $L_\ast$ for
the 10-Myr-old objects has increased on average by a factor of 4 and may reach as much as a factor of
10 for some upper mass models. The same tendency is found for the stellar radius.




\begin{table*}
\begin{centering}
\protect\protect\caption{\label{tab:6}Differences in luminosity ($\Delta L_\ast$), effective temperature
($\Delta T_{\mathrm{eff}}$) and stellar radius ($\Delta R_\ast$) between accreting
and non-accreting models in the cold accretion scenario.}
\begin{tabular}{c||ccc|}
\multirow{1}{*}{} &  & 1 Myr & \tabularnewline
\hline 
\hline 
$M_\ast$ & $\Delta L$ & $\Delta T_{\mathrm{eff}}$ & $\Delta R$\tabularnewline
{[}$M_{\odot}${]} & {[}\%{]} & {[}\%{]} & {[}\%{]}\tabularnewline
\hline 
\hline 
0.05 & -19.61 & 1.83  & -17.09\tabularnewline
0.08 & -50.27 & 1.40 & -33.67\tabularnewline
0.11 & -56.38 & 4.22 & -39.16\tabularnewline
0.13 & -57.92 & 4.48 & -40.58\tabularnewline
0.20 & -72.21 & 3.00 & -55.93\tabularnewline
0.31 & -64.93 & 1.29 & -51.12\tabularnewline
0.41 & -66.43 & 0.90 & -52.68\tabularnewline
0.56 & -48.06 & 0.62 & -28.81\tabularnewline
0.80 & -54.67 & 3.97 & -47.069\tabularnewline
1.10  & -43.74  & 5.14  & -33.12\tabularnewline
\hline 
\end{tabular}%
\begin{tabular}{ccc|}
 & 10 Myr & \tabularnewline
\hline 
\hline 
$\Delta L$ & $\Delta T_{\mathrm{eff}}$ & $\Delta R$\tabularnewline
{[}\%{]} & {[}\%{]} & {[}\%{]}\tabularnewline
\hline 
\hline 
-14.93 & -0.34 & -7.11\tabularnewline
-16.08 & -0.16 & -8.11\tabularnewline
-23.14  & 0.31 & -12.89 \tabularnewline
-17.41 & 0.79 & -10.47\tabularnewline
-22.29 & 0.15 & -11.58\tabularnewline
-22.05 & 0.24 & -12.13\tabularnewline
-29.63 & 0.24 & -16.47\tabularnewline
-4.29 & 0.82 & -3.69\tabularnewline
-9.39 & 4.68 & -13.13\tabularnewline
77.83 & 15.18 & 0.56\tabularnewline
\hline 
\end{tabular}%
\begin{tabular}{ccc|}
 & 25 Myr & \multicolumn{1}{c}{}\tabularnewline
\hline 
\hline 
$\Delta L$ & $\Delta T_{\mathrm{eff}}$ & $\Delta R$\tabularnewline
{[}\%{]} & {[}\%{]} & {[}\%{]}\tabularnewline
\hline 
\hline 
-0.17  & 0.05  & -0.12\tabularnewline
-0.65  & -0.06 & -0.19\tabularnewline
-6.05 & -0.15 & -2.78\tabularnewline
-8.31 & -0.06 & -4.13\tabularnewline
-1.24 & 0.04 & -0.11\tabularnewline
-2.03 & 0.09 & -1.20\tabularnewline
-9.15 & 0.18 & -5.03\tabularnewline
4.62 & 0.79 & 0.66 \tabularnewline
38.56 & 8.83 & -0.57\tabularnewline
-31.20 & -0.97  & -15.38\tabularnewline
\hline 
\end{tabular}
\par\end{centering}

\end{table*}

\begin{table*}

\protect\caption{\textcolor{red}{\label{tab:7}} Similar to Table~\ref{tab:2-2} but for the cold
accretion scenario.}

\begin{centering}
\textcolor{black}{}%
\begin{tabular}{c||ccc|}
\multirow{1}{*}{} &  & \textcolor{black}{1 Myr} & \tabularnewline
\hline 
\hline 
\textcolor{black}{$M_{*,\,\mathrm{fin}}$} & \textcolor{black}{$\langle\Delta L_\ast\rangle$ } & 
\textcolor{black}{$\langle\Delta T_{\rm eff}\rangle$ } & \textcolor{black}{$\langle\Delta R_\ast\rangle$}\tabularnewline
\textcolor{black}{{[}$M_{\odot}${]} } & \textcolor{black}{{[}\%{]} } & \textcolor{black}{{[}\%{]} } & \textcolor{black}{{[}\%{]}}\tabularnewline
\hline 
\hline 
\textcolor{black}{< 0.2} & \textcolor{black}{54.19 } & \textcolor{black}{2.92} & \textcolor{black}{39.62}\tabularnewline
\textcolor{black}{> 0.2} & \textcolor{black}{39.78} & \textcolor{black}{2.86} & \textcolor{black}{31.34}\tabularnewline
\hline 
\end{tabular}\textcolor{black}{}%
\begin{tabular}{ccc|}
\multicolumn{3}{c|}{\textcolor{black}{10 Myr}}\tabularnewline
\hline 
\hline 
\textcolor{black}{$\langle\Delta L_\ast\rangle$ } & \textcolor{black}{$\langle\Delta T_{\rm eff}\rangle$ } & \textcolor{black}{$\langle\Delta R_\ast\rangle$}\tabularnewline
\textcolor{black}{{[}\%{]} } & \textcolor{black}{{[}\%{]} } & \textcolor{black}{{[}\%{]}}\tabularnewline
\hline 
\hline 
\textcolor{black}{18.47} & \textcolor{black}{0.24} & \textcolor{black}{9.88}\tabularnewline
\textcolor{black}{24.75} & \textcolor{black}{4.15} & \textcolor{black}{8.24}\tabularnewline
\hline 
\end{tabular}\textcolor{black}{}%
\begin{tabular}{ccc|}
\multicolumn{3}{c}{\textcolor{black}{25 Myr}}\tabularnewline
\hline 
\hline 
\textcolor{black}{$\langle\Delta L_\ast\rangle$ } & \textcolor{black}{$\langle\Delta T_{\rm eff}\rangle$ } & \textcolor{black}{$\langle\Delta R_\ast\rangle$}\tabularnewline
\textcolor{black}{{[}\%{]} } & \textcolor{black}{{[}\%{]} } & \textcolor{black}{{[}\%{]}}\tabularnewline
\hline 
\hline 
\textcolor{black}{2.32 } & \textcolor{black}{0.06} & \textcolor{black}{1.08}\tabularnewline
\textcolor{black}{9.72} & \textcolor{black}{1.59} & \textcolor{black}{3.19}\tabularnewline
\hline 
\end{tabular}
\par\end{centering}

\textcolor{black}{\protect}
\end{table*}

\subsection{Comparison of models with different thermal efficiencies of accretion }

\begin{figure}
\begin{centering}
\resizebox{\hsize}{!}{\includegraphics{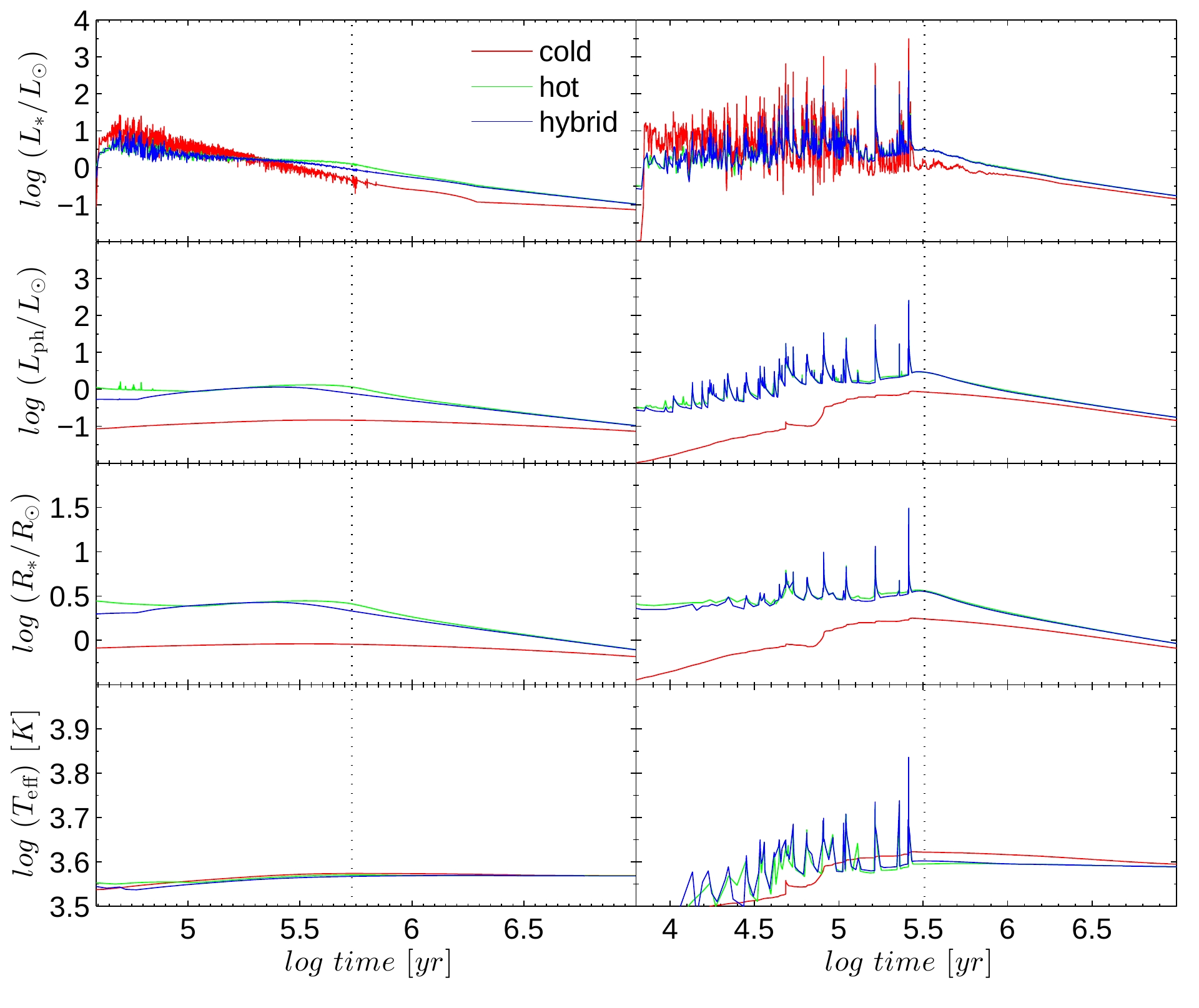}}
\par\end{centering}
\centering{}\protect\protect\protect\caption{\label{fig:9}
Time behaviour (from top to bottom) of the total luminosity, photospheric luminosity,
stellar radius, effective temperature in model 18 without accretion bursts (left column)
and model 23 with strong accretion bursts (right column). The red, green, and  blue lines
correspond to the cold, hot, and hybrid accretion scenarios. The vertical dotted lines 
define  the zero-point ages when protostars accumulate 95\% of their final mass. }
\end{figure}

To better understand the differences in the evolutionary tracks of
models with cold, hybrid and hot accretion, we plot in Figure~\ref{fig:9}
the time behavior (from top to bottom) of the total luminosity $L_\ast$,
photospheric luminosity $L_{\rm ph}$, stellar radius $R_\ast$, and effective
temperature $T_{\rm eff}$ for model 18 without strong accretion bursts (left column) 
and model 23 with strong accretion bursts (right column). The red, green, and blue lines
correspond to cold, hot, and hybrid accretion. The vertical dotted lines mark the zero-point
ages when protostars accumulate 95\% of their final mass.
The stellar radius in the hot or hybrid accretion scenarios (especially in the early 
evolutionary stages) tends to be significantly greater than that
in the cold accretion scenario, in agreement
with previous studies \citep[e.g.][]{Baraffe2012}. This trend can be explained 
by stellar bloating caused by a fraction of accretion energy absorbed by 
the star in the hybrid and hot accretion scenarios.
A similar behaviour is evident for the photospheric luminosity, which is higher in the hybrid/hot accretion
cases thanks to a larger stellar radius.
We note that the mass accretion onto the star changes the properties of the star even in the 
cold accretion case. The star becomes more compact as compared to the non-accretion case. 
Despite the accretion luminosity, the star may still be significantly below the non-accreting 
isochrone.

The behaviour of the total luminosity
(the sum of the photospheric and accretion luminosities) is somewhat more complicated. In the non-burst
case, the cold accretion model is initially characterized by a higher total luminosity than their hot
and hybrid accretion counterparts, mainly due to a lower stellar radius and, as a consequence,
a higher accretion luminosity in the cold accretion case\footnote{The stellar mass and accretion rate that enter the formula for accretion luminosity are similar in our models at a given time instance.}. In the subsequent evolution, however, the situation changes and the 
total luminosity in the hot and hybrid accretion models becomes dominant over that in 
the cold accretion model. This is due to the fact that the accretion luminosity diminishes with time
and the photospheric luminosity (which is greater in the hot and hybrid accretion models) starts to provide a major input to the total luminosity 
in the T Tauri phase \citep{Elbakyan2016}. In the burst case, the total luminosity is also on general
is smaller in the cold accretion model than in the hot and hybrid ones with one exception -- $L_\ast$
is greater in the cold accretion model during the bursts thanks to a smaller $R_\ast$ and, 
as a consequence, higher accretion luminosity.
The aforementioned trends in $L_\ast$ explain why the cold accretion models in Figure~\ref{fig:6}
demonstrate greater deviations from the non-accreting isochrones than their hot and hybrid accretion
counterparts. The systematically smaller total luminosities in models with cold accretion in the late
evolution phases result in higher $\Delta L_\ast$,  as is evident in Tables~\ref{tab:6} and \ref{tab:7}. 

The behaviour of the effective temperature is also distinct in the cold accretion and hybrid/hot accretion
models. The latter demonstrate sharp increases in $T_{\rm eff}$ during the bursts, while the former
are characterized by a rather smooth evolution of $T_{\rm eff}$. As a result, the hybrid/hot accretion
objects move to the upper-left part of the $L_\ast$--$T_{\rm eff}$ diagram during the burst, 
while the cold accretion objects -- to the upper part.
This difference in the behaviour of the cold vs.
hybrid/hot accretion models can in principle be used to constrain the accretion scenario in the protostellar
phase of evolution, provided that accretion bursts are sufficiently frequent. 
We note, however, that the hybrid/hot accretion objects can be confused with more massive stars 
in quiescence, so that care should be taken to differentiate the low-mass stars in 
the burst phase from the higher-mass stars in quiescence. 
Finally, we note that hot accretion with $\alpha\ge0.2$ fails to explain very low luminosity objects or VeLLOs \citep{Vorobyov2016}, so that a certain care should be 
taken when considering hot accretion as a viable scenario.

\section{Discussion and model caveats}

{\it Postprocessing using pre-calculated accretion rate histories.} In our models, we used the 
pre-calculated accretion rate histories derived from numerical hydrodynamics simulations of disk
formation and evolution  \citep{VB2010,VB2015}. These numerical models were shown to reproduce
the main known accretion properties of young stellar objects, e.g., the $\dot{M}$--$M_\ast$ 
steep dependence \citep{VB2008,VB2009a} and accretion burst frequency and amplitudes \citep{VB2015}.
We do not expect that self-consistent simulations coupling the stellar evolution model with the 
disk hydrodynamics 
simulations, similar to what has been done in \citet{Baraffe2016}, will significantly affect
our conclusions, because the irradiation by the central star has been taken into account
when computing the accretion rates.   
Nevertheless, we plan these self-consistent studies for the near future. 

\begin{figure*}
\begin{centering}
\includegraphics[scale=1.1]{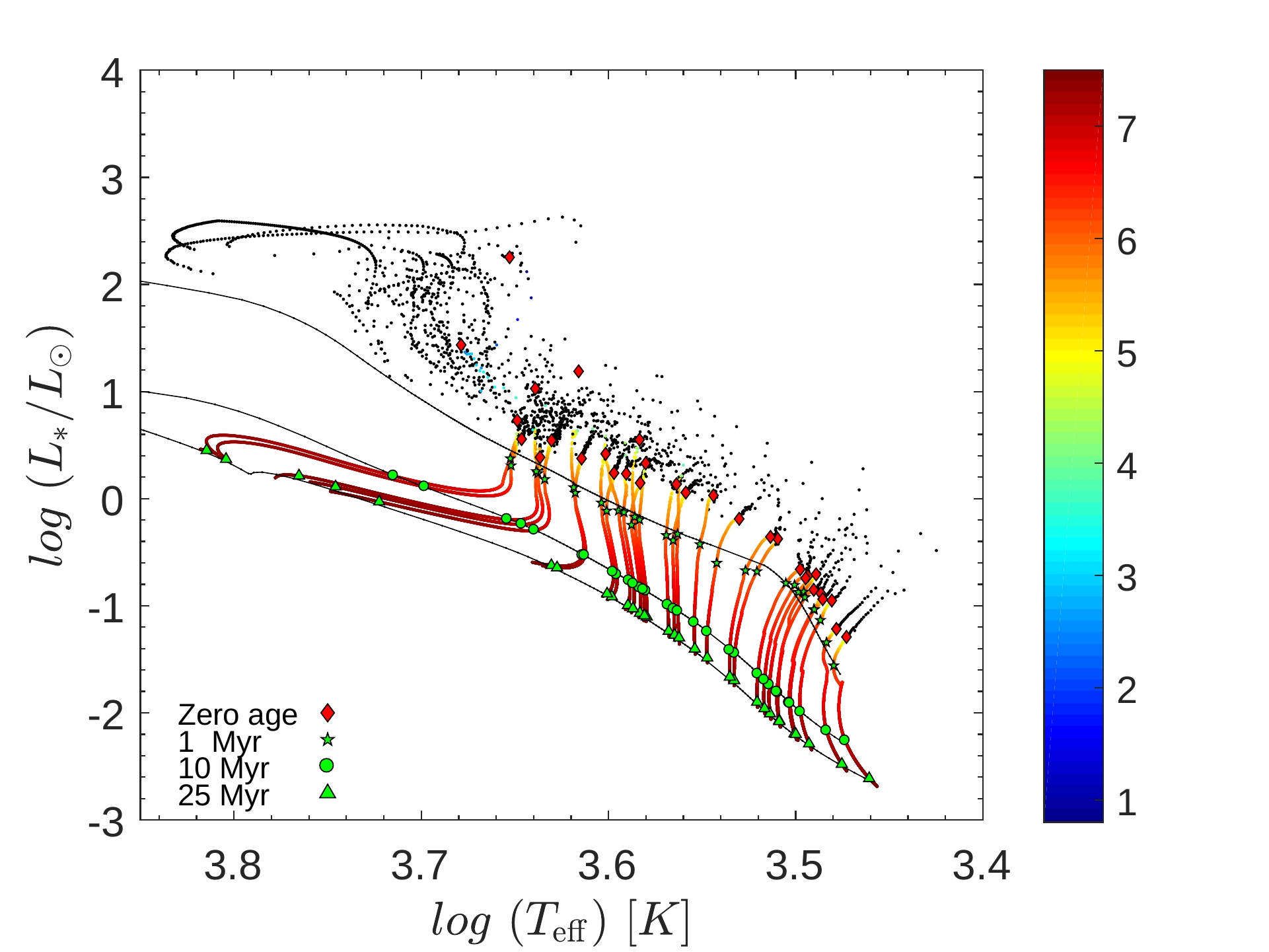} 
\par\end{centering}
\protect\protect\caption{\label{fig:10}Similar to Figure~\ref{fig:2}, but for the zero stellar age
defined as the time instance when the protostar accumulates 90\% of their final mass.}
\end{figure*}

{\it The zero stellar age.} The effect of variations in the adopted definition for
the zero stellar age has already been considered in Section~\ref{hybridaccr}. As 
Table~\ref{tab:2} demonstrates,  assuming the stellar 
zero age at the time instance when protostars accumulate 98\% of their final mass (in contrast
to the value of 95\% used throughout the paper) results in a stronger 
disagreement between the accreting models and the non-accreting isochrones. 
In the opposite case of a smaller fraction of the accreted mass, we may expect a better agreement.
Indeed, Figure~\ref{fig:10} shows the model tracks in the hybrid accretion model, but for the zero stellar
age (the red diamonds) defined as the time instance when the protostar accumulate 90\% of their final
mass. This definition of the zero age corresponds to the evolutionary phase when 10\% of the
final stellar mass is still confined in the disk and residual envelope, which is more appropriate
to the late Class I phase of protostellar evolution rather than to the T Tauri phase. As Figure~
\ref{fig:1} shows, accretion rates at these times may still show some substantial variability.
A visual comparison of Figures~\ref{fig:2} and \ref{fig:10} demonstrates that the agreement
of the 1.0-Myr-old accreting objects (the green stars)  
with the corresponding non-accreting isochrone is now better than in the case 95\% case. However,
the zero age (or the birth) locations are now widely scattered, because the protostars continue 
accreting at this stage and may experience significant excursions in the value of $L_\ast$ due to
the time-varying accretion rates in the late protostellar phase. Time variations
in the birth locations of stars present no problem and may be a real phenomenon as was noted in
\citet{Baraffe2012}. The classic smooth birthline of stars introduced in \citet{Stahler1983}
was calculated based on spherically symmetric collapse simulations adopting 
a constant protostellar accretion rate and neglecting the possible significant effect 
of accretion variability introduced by the circumstellar disk\footnote{In fact, the very existence
of the disk was questioned in \citet{Stahler1983}.}.  However, the 90\% definition 
may still technically correspond to the  late protostellar phase (rather than T Tauri phase) 
and we therefore think that taking 
the 90\% case as the zero age point is less physically motivated than the other two cases. 


{\it The thermal efficiency of accretion.} In this work, we have adopted $\alpha=0.1$ for 
the maximum fraction of absorbed energy by the central object 
and $\dot{M}_{\rm cr}=10^{-5}~M_\odot$~yr$^{-1}$ for
the transition from cold to hot accretion (see Section~\ref{Ascenarios}).  
The possible variations in these quantities may affect somewhat the stellar tracks on
the $L_\ast$--$T_{\rm eff}$ diagram. For instance, increasing the value of $\dot{M}_{\rm cr}$ 
we will simultaneously decrease the time during which the star may absorb  part of the accretion
energy. In the limit of $\dot{M}_{\rm cr}\gg 10^{-4}~M_\odot$~yr$^{-1}$, the hybrid accretion 
tracks will behave similar to those of the cold accretion scenario, 
because accretion bursts of such high magnitude are extremely rare.
In the opposite case of $\dot{M}_{\rm cr}\ll 10^{-6}~M_\odot$~yr$^{-1}$, the hybrid accretion 
track will resemble those of the hot accretion scenario, because the star will spend most of its 
time absorbing part of the accretion energy.  In addition, varying the value of $\alpha$ may produce a 
spread in the $L_\ast$--$T_{\rm eff}$ diagram, as was shown in \citet{Baraffe2012}. In the
hot accretion scenario, however, the values of $\alpha$ greater than 0.2 are unlikely, because the resulting
photospheric luminosities are always greater than $0.1~L_\odot$. This contradicts the 
apparent existence of very low luminosity objects with the internal luminosity $< 0.1~L_\odot$ 
in the protostellar phase \citep[see][]{Vorobyov2016}.

{\it Initial stellar radius and mass.}  To check the effect of the initial conditions 
imposed on the protostellar seed, we re-calculated some typical models in the 
hot and cold accretion scenarios by varying the initial seed radius from 2.7 to 4.0 Jovian radii and reducing the initial seed mass to 4.0 Jovian masses. 
The resulting pre-main-sequence evolution showed no significant deviation from the evolution of the
original models. Unfortunately, we were not able to vary the initial seed parameters over a wider 
range of radii and masses due to numerical divergence of the stellar evolution code. 
We plan to perform a more rigorous study
of the effects of initial conditions in the future self-consistent simulations.

\section{Conclusions}
In this paper, we have considered the pre-main-sequence evolution of low-mass stars and  brown dwarfs
starting from the formation of a protostellar seed and taking into account the mass accretion 
during the initial several Myr of evolution. The stellar evolution was computed
using the STELLAR evolution code originally developed by \citet{Yorke2008} and further modified by
\citet{Hosokawa2013}.

The mass accretion rates were taken from numerical hydrodynamics simulations of disk evolution 
starting from the gravitational collapse of
pre-stellar cloud cores of various mass and angular momentum. The resulting accretion rates
exhibit various patterns of time behavior: from smoothly declining rates to strongly time-varying ones featuring episodic bursts similar in magnitude to those of the FU-Orionis-type eruptions \citep{VB2010,VB2015}.
Three scenarios for the thermal efficiency of accretion were considered: hot accretion 
with a constant fraction
of accretion energy absorbed by the central object (star or brown dwarf), cold accretion in which essentially
all accreted energy is radiated away, and hybrid accretion in which the fraction of absorbed 
energy depends on the accretion rate. We compare the resulting stellar evolution tracks in the total
luminosity ($L_\ast$) -- effective temperature ($T_{\rm eff}$) diagram with the isochrones and isomasses
derived using the non-accreting models of Yorke \& Bodenheimer. Our key findings can be summarized 
as follows.

\begin{itemize}

\item In the hybrid accretion case, young 1.0-Myr-old objects show notable deviations from the non-accreting
isochrones and isomasses for both low-mass stars and brown dwarfs. The largest deviations (relative
to the non-accreting models) are found for the total stellar luminosity and stellar radius (19\%--51\%
and 8\%--32\%, respectively), while for the effective temperature the deviations are relatively 
mild (0.15\%--2.1\%). The disagreement between the accreting and non-accreting models diminishes with stellar age, remaining within several per cent for 10-Myr-old and 25-old-old objects. The calculated deviations depend somewhat on our definition of the stellar zero age.

\item The hot accretion case is qualitatively similar to hybrid accretion, but showing
somewhat smaller deviations for $L_\ast$ and $R_\ast$. 

\item The cold accretion case features the largest deviations from the non-accreting models of Yorke
\& Bodenheimer. For the  1.0-Myr-old objects, the deviations in $L_\ast$ in $R_\ast$ 
are on average factors of 2--3 greater than in the cases of hybrid/hot accretion. A disagreement
is also found for the 10-Myr-old objects and even for the 25-Myr-old objects, especially for the upper
mass ones. The greater disagreement as compared to the hybrid/hot accretion cases 
can be explained by the systematically smaller stellar radii and, as consequence, 
photospheric luminosities (which dominate
the accretion luminosity in the T Tauri phase \citep{Elbakyan2016}) in models with cold accretion.

\item As a result of this mismatch, the use of the
$L_\ast$--$T_{\rm eff}$ diagram may lead to the false age estimate for 
objects with $T_{\rm eff}>3500$~K, as was also previously noted in \citet{Baraffe2009} and \citet{Hosokawa2011}.
For instance, 1.0-Myr-old objects with cold accretion can be falsely identified as 4.5-Myr-old ones.
For the case of hybrid and hot accretion, the error remains within a factor of 2 for
object of 1.0 Myr age, but diminishes for older objects.

\item Hybrid and hot accretion models show sharp increases
in both $L_\ast$ and $T_{\rm eff}$ during accretion bursts. As a result, these models show 
notable excursions  to the upper-left region of the $L_\ast$--$T_{\rm eff}$ diagram.
On the contrary, $T_{\rm eff}$ in the cold accretion scenario is weakly affected by the 
bursts and the cold accretion objects show strong surges only in $L_\ast$.
These differences between the
stellar evolution tracks of the hybrid/hot accretion and cold accretion objects experiencing accretion
bursts can potentially be used to constrain the thermal efficiency of accretion, but care 
should be taken in order
not to confuse the low-mass stars in the burst phase with the upper-mass stars in quiescence.

\item
The increase in luminosity of the cold accretion objects experiencing accretion bursts
is notably stronger than that of the hybrid/hot accretion objects. In particular, the cold accretion
objects sometimes exceed 1000~$L_\odot$ in luminosity, whereas the hybrid/hot accretion objects 
never exceed 400~$L_\odot$ in luminosity. 
According to the recent review by \citet{Audard2014},
the strongest luminosity outbursts are observed in FU~Ori (340-500~$L_\odot$), Z~CMa (400-600~$L_\odot$),
and V1057 (250-800~$L_\odot$), all of which are below 1000~$L_\odot$. 
Therefore, the lack of very strong luminosity outbursts in the FU-Orionis-type objects 
may be an indirect evidence in favour of the hybrid/hot scenarios for the thermal efficiency of accretion.
The hot scenario with $\alpha\ge 0.2$, however, fails to reproduce the very low 
luminosity objects (VeLLOs) due to the fact that the resulting photospheric luminosity 
(which sets the floor for the total luminosity in the early stages of evolution) is always greater 
than the VeLLO limit of 0.1~$L_\odot$ \citep{Vorobyov2016}. 

\end{itemize}
In the Appendix, we provide the updated isochrones based on the stellar evolution models taking mass accretion into account. The utility of these isochrones hinges on the adopted scenario for the thermal efficiency
of accretion and efforts
now should be placed to find the means of differentiating between the hybrid/hot and cold accretion
cases,
similar to what have recently been done in \citet{Geroux2016}.

\section{Acknowledgements}
We are thankful to the anonymous referee for helpful suggestions that allowed us to 
to improve the manuscript.
E.I.V. and V.G.E are thankful to Isabelle Baraffe and Gilles Chabrier for stimulating discussions that
inspired this work.
E. I. Vorobyov and V.G. Elbakyan acknowledges supported by the Russian Ministry of Education and Science Grant 3.5602.2017. V.G. Elbakyan acknowledges the Southern Federal University for 
financial support with the international travel grant.
The simulations were performed on the Vienna Scientific Cluster (VSC-2 and VSC-3) and on the
Shared Hierarchical Academic Research Computing Network (SHARCNET).
We also appreciate the financial supports by the Grants-in-Aid for Basic Research
by the Ministry of Education, Science and Culture of Japan 
(16H05996: TH) and by Grant-in-Aid for JSPS Fellows (SH).
Portions of this work were conducted at the Jet Propulsion Laboratory,
California Institute of Technology, operating under a contract with 
the National Aeronautics and Space Administration (NASA).

\begin{appendix}

\section{Updated isochrones using accreting models}

\begin{table*}
\protect\caption{\label{tab:5}Polynomial coefficients used for calculating the accretion-model
isochrones}
\begin{centering}
\begin{tabular}{c||cccc}
\hline 
 & Cold & Cold & Hot & Hybrid\tabularnewline
 & 1 Myr & 10 Myr & 1 Myr & 1 Myr\tabularnewline
\hline 
\hline 
$p_{1}$  & $5.298\times10^{2}$  & $2.665\times10^{2}$  & $-7.249\times10^{3}$  & $0.9761\times10^{2}$\tabularnewline
$p_{2}$  & $-7.582\times10^{3}$  & $-3.858\times10^{3}$  & $1.039\times10^{5}$  & $-1.050\times10^{3}$\tabularnewline
$p_{3}$  & $4.067\times10^{4}$  & $2.092\times10^{4}$  & $-5.580\times10^{5}$  & $3.770\times10^{3}$\tabularnewline
$p_{4}$  & $-9.690\times10^{4}$  & $-5.033\times10^{4}$  & $1.332\times10^{6}$  & $-4.512\times10^{3}$\tabularnewline
$p_{5}$  & $8.652\times10^{4}$  & $4.534\times10^{4}$  & $-1.193\times10^{6}$  & $-1.034\times10^{-8}$\tabularnewline
\hline 
\end{tabular}
\par\end{centering}
\end{table*}

\begin{figure}
\begin{centering}
\resizebox{\hsize}{!}{\includegraphics{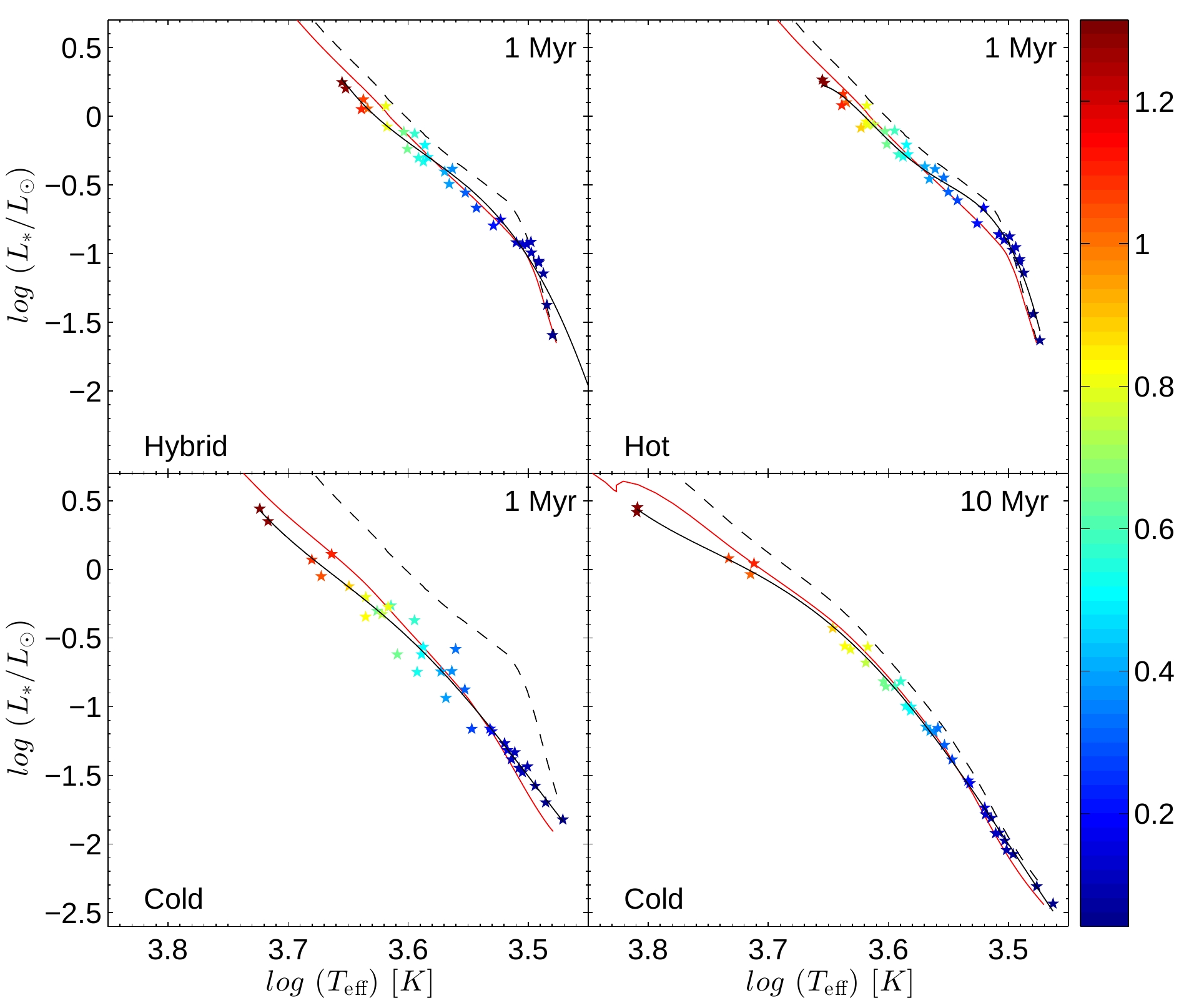}}
\par\end{centering}
\protect\protect\protect\caption{\label{fig:8}
Total luminosity--effective temperature diagram at 1.0~Myr and 10~Myr in the  hybrid, 
hot and cold accretion scenarios. The coloured symbols represent our accreting models
and the color of the symbols
varies according to the stellar mass shown in the vertical bar (in $M_{\odot}$). 
The corresponding ages of the model data and scenarios are indicated in each panel.
The black dashed lines are the isochrones (of the corresponding age)
derived from the non-accreting models of \citet{Yorke2008}, while the black solid lines
are the best-fit curves to our model data (accreting isochrones, see the text). 
The red solid lines represent the
non-accreting isochrones of Yorke \& Bodenheimer which fit best to our model data. }
\end{figure}

In this section, we calculate the isochrones taking accretion into account and compare them with the
isochrones derived from the non-accreting models of \citet{Yorke2008}.
The coloured symbols in Figure~\ref{fig:8} represent our accreting models in 
the $L_\ast$--$T_{\rm eff}$ diagram at 1.0~Myr and 10~Myr in the  hybrid, hot and cold accretion
 scenarios. The corresponding ages
and scenarios are indicated in the panels. The color of the symbols
varies according to the stellar mass shown in the vertical bar (in $M_{\odot}$). 
The black dashed lines are the corresponding isochrones 
derived from the non-accreting models of \citet{Yorke2008}, while the black solid lines
are the best-fit curves to our model data (hereafter, accreting isochrones). 
The interpolation is done using the Vandermonde matrix and a forth-order polynomial of the form
\begin{equation}
 \log {L_\ast \over L_\odot} =p_{1}(\log T_{\rm eff})^{4}+p_{2}(\log T_{\rm eff})^{3}+p_{3} 
 (\log T_{\rm eff})^{2}+p_{4}\log T_{\rm eff}+p_{5},
\end{equation}
and the coefficients of the polynomical are listed in Table~\ref{tab:5}.

Evidently, there exists a notable mismatch between the non-accreting (black dashed lines) 
and accreting (black solid lines) isochrones, especially for the cold accretion objects of
1~Myr age. To quantify this disagreement, we plot in Figure~\ref{fig:8} with the red solid lines 
the non-accreting isochrones which best fit our model data.  The resulting ages of these isochrones
are: 1.5~Myr (top-left panel), 1.5~Myr (top-right panel), 4.5~Myr (bottom-left panel), and 16~Myr 
(bottom-right panel).  As was discussed earlier, the largest error in the age determination 
when using the non-accreting isochrones can occur for young, intermediate and upper-mass 
objects with cold accretion --
the 1.0-Myr-old objects can be falsely identified as 4.5-Myr-old ones. The error
still remains significant at 10~Myr -- these objects can be falsely interpreted 
as 16-Myr-old ones.  For the case of hybrid and hot accretion, the error 
remains within a factor of 2 for object of 1.0~Myr age, but diminishes for older
objects. 

In order for the accreting isochrones to be of practical use, efforts 
should now be focused on determining the most plausible scenario for the thermal efficiency of accretion.
Multidimensional numerical hydrodynamics simulations of accreting stars, similar to what have recently
been done by \citet{Geroux2016}, can give us a clue about the thermal efficiency of accretion, 
but usually these simulations are too computationally
expensive to be used for simulating the long-term accretion history of (sub-)solar mass stars.
Nevertheless, they can be useful for calibrating the less computationally expensive 
one-dimensional models of stellar evolution.



\end{appendix}


\begin{thebibliography}{}

\bibitem[\protect\citeauthoryear{Audard et al.}{2014}]{Audard2014}
Audard, M., \'Abrah\'am, P., Dunham, M. M., et al. 2014, in Protostars and
Planets VI, ed. H. Beuther, R. S. Klessen, C. P. Dullemond, \& T. Henning
(Tucson, AZ: Univ. Arizona Press), 387

\bibitem[\protect\citeauthoryear{Baraffe et al.}{1998}]{Baraffe1998}
Baraffe, I., Chabrier, G., Allard, F., \& Hauschildt, P. H. 1998, A\&A,
337, 403

\bibitem[\protect\citeauthoryear{Baraffe et al.}{2009}]{Baraffe2009}
Baraffe, I., Chabrier, G., \& Gallardo, J. 2009, ApJ, 702, L27 

\bibitem[\protect\citeauthoryear{Baraffe et al.}{2012}]{Baraffe2012}
Baraffe, I., Vorobyov, E. I., \& Chabrier, G. 2012, ApJ, 756, 118

\bibitem[\protect\citeauthoryear{Baraffe et al.}{2016}]{Baraffe2016}
Baraffe, I., Elbakyan, V. G., Vorobyov, E. I., \& Chabrier, G. 2016, A\&A, in press


\bibitem[\protect\citeauthoryear{D'Antona \& Mazzitelli}{1994}]{DAM}
D'Antona, F., \& Mazzitelli, I. 1994, ApJS, 90, 467

\bibitem[\protect\citeauthoryear{Dunham et al.}{2008}]{Dunham2008}
Dunham, M. M., Crapsi, A. Evans N. J. II, et al. 2008,
ApJSS, 179, 249

\bibitem[\protect\citeauthoryear{Dunham et al.}{2010}]{Dunham2010}
Dunham, M. M., Evans, N. J. II, Terebey, S., Dullemond, C. P., Young, C. H.  2010, ApJ, 710, 470

\bibitem[\protect\citeauthoryear{Dunham et al.}{2014}]{Dunham2014}
Dunham, M. M., Vorobyov, E. I., Arce, G. H. 2014, MNRAS, 444, 887

\bibitem[\protect\citeauthoryear{Elbakyan et al.}{2016}]{Elbakyan2016}
Elbakyan, V. G., Vorobyov, E. I., Glebova, G. M. 2016, Astron. Rep., 60, 879

\bibitem[\protect\citeauthoryear{Geroux et al.}{2016}]{Geroux2016}
Geroux, C., Baraffe, I., Viallet, M., et al. 2016, A\&A, 588, 85

\bibitem[\protect\citeauthoryear{Hartmann et al.}{1997}]{Hartmann1997}
Hartmann, L., Cassen, P., \& Kenyon, S. J. 1997, ApJ, 475, 770

\bibitem[\protect\citeauthoryear{Hartmann et al.}{2011}]{Hartmann2011}
Hartmann, L., Zhu, Z., \& Calvet, N. 2011, arXiv:1106.3343

\bibitem[\protect\citeauthoryear{Hartmann et al.}{2016}]{Hartmann2016}
Hartmann, L., Herczeg, G., \& Calvet, N. 2016, ARA\&A, 54, 135

\bibitem[\protect\citeauthoryear{Hayashi}{1961}]{Hayashi1961}
Hayashi, C. 1961, PASJ, 13, 450

\bibitem[\protect\citeauthoryear{Henyey et al.}{1955}]{Henyey1955}
Henyey, L.G., Lelevier, R. \& Levee, R. D. 1955, PASP, 67, 154

\bibitem[\protect\citeauthoryear{Hosokawa \& Omukai}{2009}]{Hosokawa2009}
Hosokawa, T., \& Omukai, K. 2009, ApJ, 691, 823 

\bibitem[\protect\citeauthoryear{Hosokawa et al.}{2011}]{Hosokawa2011}
Hosokawa, T., Offner, S., \& Krumholz, M. 2011, ApJ, 738, 140 

\bibitem[\protect\citeauthoryear{Hosokawa et al.}{2013}]{Hosokawa2013}
Hosokawa, T., Yorke, H., Inayoshi, K.  et al. 2013, ApJ, 778, 178

\bibitem[\protect\citeauthoryear{Hosokawa et al.}{2016}]{Hosokawa2016}
Hosokawa, T., Hirano, S., Kuiper, R., et al. 2016, ApJ, 824, 119

\bibitem[\protect\citeauthoryear{Kley \& Lin}{1996}]{Kley1996}
Kley, W. \& Lin, D. N. C. 1996, ApJ, 461, 933

\bibitem[\protect\citeauthoryear{Kunitimo et al.}{2017}]{Kunitomo2017}
Kunitomo, M., Guillot, T., Takeuchi, T., \& Ida, S.  2017, A\&A, 599, A49 

\bibitem[\protect\citeauthoryear{Kuiper \& Yorke}{2013}]{KY13}
Kuiper, R., \& Yorke, H.~W. 2013, ApJ, 772, 61

\bibitem[\protect\citeauthoryear{Liu et al.}{2016}]{Liu2016}
Liu, H. B., Takami, M. \& Kudo, T. et al. 2016, Science Advances, 200875

\bibitem[\protect\citeauthoryear{Machida et al.}{2011}]{Machida2011}
Machida, M. N., Inutsuka, S., \& Matsumoto, T. 2011, ApJ, 729, 42

\bibitem[\protect\citeauthoryear{Palla \& Stahler}{1990}]{Palla1990}
Palla, F., \& Stahler, S. W. 1990, ApJ, 360, 47

\bibitem[\protect\citeauthoryear{Palla \& Stahler}{1991}]{Palla1991}
Palla, F., \& Stahler, S. W. 1991, ApJ, 375, 288 

\bibitem[\protect\citeauthoryear{Palla \& Stahler}{2000}]{Palla2000}
Palla, F., \& Stahler, S. W. 2000, ApJ, 540, 255

\bibitem[\protect\citeauthoryear{Palla \& Stahler}{2000}]{Pascucci2016}
Pascucci, I., Testi, L., Herczeg, G. J., et al. 2016, ApJ, 831, 125

\bibitem[\protect\citeauthoryear{Sakurai et al.}{2015}]{Sakurai2015}
Sakurai, Y., Hosokawa, T., Yoshida, N., Yorke, H. W. 2015, MNRAS, 452, 755

\bibitem[\protect\citeauthoryear{Shu}{1977}]{Shu1977}
Shu, F. S. 1977, ApJ, 214, 488

\bibitem[\protect\citeauthoryear{Soderblom et al.}{2014}]{Soderblom2014}
Soderblom, D. R., Hillenbrand, L. A., Jeffries, R. D. et al. 2014, Protostars \& Planets VI,
H. Beuther, R. S. Klessen, C. P. Dullemond, and T. Henning (eds.), 
University of Arizona Press, Tucson, 914, 219

\bibitem[\protect\citeauthoryear{Stahler et al.}{1980}]{Stahler1980}
Stahler, S. W., Shu, F., \& Taam, R. E. 1980, ApJ, 241, 637

\bibitem[\protect\citeauthoryear{Stahler}{1983}]{Stahler1983}
Stahler, S. W. 1983, ApJ, 274, 822

\bibitem[\protect\citeauthoryear{Tomida et al.}{2017}]{Tomida2017}
Tomida, K., Machida, M. N., Hosokawa, T. et al. 2017, ApJ, 835, L11

\bibitem[\protect\citeauthoryear{Tsukamoto et al.}{2016}]{Tsukamoto2016}
Tsukamoto, Y., Okuzumi, S., Kataoka, A. 2016, ApJ, in press

\bibitem[\protect\citeauthoryear{Vorobyov \& Basu}{2006}]{VB2006}
Vorobyov, E. I., \& Basu, S., 2006, ApJ, 650, 956

\bibitem[\protect\citeauthoryear{Vorobyov \& Basu}{2008}]{VB2008}
Vorobyov, E. I., \& Basu, S., 2008, ApJL, 676, 139

\bibitem[\protect\citeauthoryear{Vorobyov \& Basu}{2009a}]{VB2009a}
Vorobyov, E. I., \& Basu, S., 2009, ApJ, 703, 922

\bibitem[\protect\citeauthoryear{Vorobyov \& Basu}{2009b}]{VB2009b}
Vorobyov, E. I., \& Basu, S., 2009, MNRAS, 393, 822

\bibitem[\protect\citeauthoryear{Vorobyov}{2010}]{Vorobyov2010}
Vorobyov, E. I., 2010, ApJ, 723, 1294

\bibitem[\protect\citeauthoryear{Vorobyov \& Basu}{2010}]{VB2010}
Vorobyov, E. I., \& Basu, S. 2010, ApJ, 719, 1896

\bibitem[\protect\citeauthoryear{Vorobyov \& Basu}{2015}]{VB2015}
Vorobyov, E. I., \& Basu, S., 2015, ApJ, 805, 115

\bibitem[\protect\citeauthoryear{Vorobyov et al.}{2017}]{Vorobyov2016}
Vorobyov, E. I., Elbakyan, V. G., M. M. Dunham, Guedel, M. 2016, A\&A, 600, 36

\bibitem[\protect\citeauthoryear{Yorke \& Bodenheimer}{2008}]{Yorke2008}
Yorke H. W., \& Bodenheimer P., 2008, in Beuther H., Linz H., Henning T.,
eds, ASP Conf. Ser. Vol. 387, Massive Star Formation: Observations
Confront Theory. Astron. Soc. Pac., San Francisco, p. 189

\end{thebibliography}
\end{document}